\documentclass[apj]{emulateapj}
\usepackage{graphicx,psfig,amssymb,float,natbib,rotating,txfonts,lscape,rotating}
\bibpunct{(}{)}{;}{a}{}{,}

\newcommand{\HST}{{\em HST}}

\begin{document}

\title{{\em ChaMPlane} Deep Galactic Bulge Survey. I. Faint
accretion-driven binaries in the Limiting Window}
\shorttitle{Faint
accretion-driven binaries in the Limiting Window}
\shortauthors{van den Berg, Hong, and Grindlay}

\author{Maureen van den Berg\altaffilmark{1}, JaeSub
Hong\altaffilmark{1}, Jonathan E. Grindlay\altaffilmark{1}} 
\altaffiltext{1}{Harvard-Smithsonian Center for Astrophysics, 60
Garden Street, Cambridge, MA 02138, USA; maureen@head.cfa.harvard.edu}

\begin{abstract}
We have carried out a deep X-ray and optical survey with {\em Chandra}
and {\em HST} of low-extinction regions in the Galactic bulge.  Here
we present the results of a search for low-luminosity
($L_X\lesssim10^{34}$ ergs s$^{-1}$) accreting binaries among the {\em
Chandra} sources in the region closest to the Galactic Center, at an
angular offset of 1.4$^{\circ}$, that we have named the Limiting
Window. Based on their blue optical colors, excess H$\alpha$ fluxes,
and high X-ray--to--optical flux ratios, we identify three likely
accreting binaries; these are probably white dwarfs accreting from
low-mass companions (cataclysmic variables or CVs) although we cannot
exclude that they are quiescent neutron-star or black-hole low-mass
X-ray binaries. Distance estimates put these systems farther than
$\gtrsim$2 kpc.  Based on their H$\alpha$-excess fluxes and/or high
X-ray--to--optical flux ratios, we find 22 candidate accreting
binaries; however, the properties of some can also be explained if
they are dMe stars or active galaxies. We investigate the CV number
density towards the bulge and find that the number of observed
candidate CVs is consistent with or lower than the number expected for
a constant CV-to-star ratio that is fixed to the local value.  Our
conclusions are limited by uncertainties in the extinction (for which
we see a $\sim$30\% variation in our
6.6\arcmin$\times$6.6\arcmin~field) and selection effects. The X-ray
properties of two likely CVs are similar to those of the faint, hard
X-ray sources in the Galactic-Center region that have been explained
by (mainly) magnetic CVs. If our candidates belong to the same
population, they would be the first members to be optically
identified; optical or infrared identification of their
Galactic-Center analogs would be impossible due to the higher
obscuration. We speculate that all Galactic hard X-ray sources in our
field can be explained by magnetic CVs.
\end{abstract}


\section{Introduction}

With its sharp view, the {\em Chandra} X-ray Observatory has revealed
a large population of X-ray point sources along the line of sight
towards the central 40 pc of our Galaxy \citep{munoea03}. The majority
of these are faint ($L_X\approx10^{31-34}$ ergs s$^{-1}$) and have
intrinsically hard X-ray spectra. Comparison with the X-ray properties
of candidate local analogs suggests that the Galactic-Center sources
are a mix of compact objects (white dwarfs, neutron stars and black
holes) accreting from binary companions. From the small fraction of
sources with bright ($K<15$) infrared candidate counterparts,
\cite{laycea05} excluded that the contribution of neutron stars in
wind-fed Be high-mass X-ray binaries (HMXBs) exceeds $\sim$10\%; on
the other hand, a significant contribution from wind-fed HMXBs to
X-ray sources in the central 2$^{\circ}$ $\times$ 0.8$^{\circ}$ had
been suggested by \cite{pfahea02}.  This leaves magnetic white dwarfs
accreting from low-mass companions (i.e.~magnetic cataclysmic
variables or CVs) as the most likely dominating source class, as had
been suggested by \cite{munoarabea04}. However, this also leaves us
with the enormous challenge to identify these sources at other
wavelengths: CVs in the Galactic Center are too faint to be detected
in the infrared (unless they undergo a nova or bright dwarf-nova
outburst), and the Galactic-Center region is too heavily obscured by
dust to search for counterparts in the optical. At the same time, by
the sheer number of Galactic-Center X-ray sources (thousands), this
population has great importance for understanding the Galactic
accreting binary content and unraveling the evolutionary history of
the inner Galaxy.

As part of the {\em Chandra} Multi-wavelength Plane survey \citep[{\em
ChaMPlane};][]{grinhongea05}, we have carried out an X-ray and optical
survey of deep, near-simultaneous {\em Chandra} and {\em HST}
pointings of three low-extinction regions, or {\em Windows}, in the
Galactic bulge.  These Windows lie on a nearly radial alignment with
the Galactic Center, at angular offsets between 1.4$^{\circ}$ and
4$^{\circ}$. One of the aims of this survey is to look for signatures
of a population of faint, hard X-ray sources similar to the one found
by \cite{munoea03}, and take advantage of the more favorable
conditions for source identification compared to the Galactic
Center. Indeed, based on the spatial distribution of {\em Chandra}
sources in these three and four more inner-bulge fields,
\cite{hongea09} (H09) show that this population can be traced out to
at least 1.4$^{\circ}$ from the Galactic Center, i.e.~out to our
``innermost'' Window.  We have named this field the Limiting Window
(LW) because extinction rapidly increases (and therefore chances for
optical identification drastically decline) when moving closer to the
Galactic Center. In this paper we try to identify X-ray binaries (that
will be most likely CVs given their relative abundance) among {\em
Chandra} sources in the LW by searching for optical counterparts; by
extrapolation, this could also shed light on the nature of the
Galactic-Center sources.

Another main goal of our Windows survey is to study for the first time
the population of CVs towards the bulge.  Much of what we know about
the Galactic CV population is based on a sample of relatively local
systems, mainly within $\sim$2 kpc from the Sun (see, for example,
\cite{warn95}, \cite{akea08}). Our deep {\em Chandra} exposures have
the sensitivity to detect bright ($L_X\gtrsim10^{31}$ ergs s$^{-1}$)
CVs at $\gtrsim$8 kpc, which means we can significantly increase the
depth of existing CV samples.  In a follow-up paper we will combine
our results for the LW with the search for CVs and other accreting
binaries in our other two Windows---Baade's and Stanek's Windows
(initial results were already given in \cite{vdbergea06}).  Limits on
the CV space density towards the Galactic anti-Center were derived by
\cite{grinhongea05} from {\em ChaMPlane} data.

After describing the X-ray and optical observations in
\S\ref{sec_obs}, we outline the analysis steps 
in \S\ref{sec_ana}. We present the identified candidate CVs in
\S\ref{sec_res}, and discuss our results in the contexts of the
Galactic CV population and the Galactic-Center X-ray--source
population in \S\ref{sec_dis}. We end with our conclusions in
\S\ref{sec_con}.

\section{Observations} \label{sec_obs}

\subsection{Chandra data}

The Limiting Window was observed for 100 ks with the {\em
Chandra}/ACIS-I aimpoint placed at Galactic coordinates
($l$,$b$)=(0.10$^{\circ}$,$-$1.43$^{\circ}$).  Due to observatory
constraints, the observation was split in three exposures: ObsID 6362
(2005 Aug 19, 38 ks), ObsID 5934 (2005 Aug 22, 41 ks) and ObsID 6365
(2005 Oct 25, 21 ks). We stacked the three images into a single image;
in the remainder of the paper we only consider the stacked
pointing. The data were processed with the {\em ChaMPlane} X-ray
data-reduction pipeline \citep{hongvandea05}. We performed source
detection on images in a soft (0.3--2.5 keV), hard (2.5--8 keV) and
broad ($B_x=0.3$--8 keV) energy band, and merged the resulting source
lists into a master catalog following the procedures described in H09.
A total of 319 ACIS-I sources are detected. The pipeline products that
we mainly use in this paper are source positions and corresponding
95\% confidence radii $r_{95}$, net source counts in the $B_x$ band
extracted from a region enclosing 95\% of the energy at 1.5 keV, and
energy quantiles $E_x$ denoting the energies below which $x$\% of the
counts in the $B_x$ band are detected. We use the latter to constrain
X-ray spectral parameters for low-count sources; see
\cite{hongschlea04} for a description of the quantile method. The
complete {\em Chandra} source list can be found in H09.

\subsection{HST/ACS data}

We observed with \HST~the inner area of the ACIS field with a
2$\times$2 mosaic of slightly overlapping pointings of the Wide Field
Camera (WFC) on the Advanced Camera for Surveys (ACS). The
observations (program GO-10353) were carried out simultaneous with the
first {\em Chandra} exposure (2005 Aug 19). A single WFC pointing
images a 3.4\arcmin$\times$3.4\arcmin~field with
$\sim$0.05\arcsec~pixels using two CCD detectors separated by a
2.5\arcsec~gap.  Exposures were taken through the F435W
(``$B_{435}$''), F625W (``$R_{625}$'', similar to Sloan $r$) and F658N
(H$\alpha$) filters.  Each tile of the mosaic was observed with the
same exposure sequence: 4$\times$492 s in F435W, 168 s + 2$\times$167
s in F625W and 4$\times$496 s + 4$\times$492 s in F658N. The offsets
in Galactic coordinates ($\Delta l, \Delta b$) of each {\em HST}
pointing from the center of the {\em Chandra} pointing are
($-$0.04$^{\circ}$,+0.01$^{\circ}$),
(+0.04$^{\circ}$,$-$0.01$^{\circ}$), (+0.01$^{\circ}$,+0.04$^{\circ}$)
and ($-$0.01$^{\circ}$,$-$0.04$^{\circ}$).  No dithering was applied
to fill in the WFC chip gap. After boresight correction (see
\S\ref{sec_oid}), 100 {\em Chandra} sources are included in the field
of view of the ACS mosaic, including one source whose 95\% X-ray error
circle lies partially in the chip gap. The sources that are covered by
our ACS mosaic are listed in Table~\ref{tab_srclist}.

\begin{table}
\begin{center}
\caption{{\em Chandra} sources in the ACS field of view \label{tab_srclist}}
\begin{tabular}{lrl}
\hline
\hline
CXOPS\,J & $S/N_{H_c}$ & ID (table) \\
\hline
175139.2$-$293434 & $-$0.51  & \ldots                      \\ 
175138.7$-$293519 &   2.83   & \ldots                      \\ 
175137.5$-$293602 &   0.42   & LW\,28 (\ref{tab_otherid})  \\ 
175137.4$-$293515 &   6.74   & LW\,41 (\ref{tab_otherid})  \\ 
175133.6$-$293313 &   6.38   & LW\,19 (\ref{tab_candfxfo}) \\ 
\hline
\end{tabular}
\end{center}

For each {\em Chandra} source in the field of the ACS mosaic we list
the source name and signal-to-noise ratio in the hard band ($H_c$=2--8
keV). If appropriate, we also give the short name adopted in this
paper and (in parentheses) the number of the table with the X-ray and
optical properties. A negative value of $S/N_{H_c}$ indicates that the
source was not significantly detected in the $H_c$ band. {\em The
complete table is available in the electronic edition of the journal,
here we only list the first lines to illustrate the table format and
content.}
\end{table}

\section{Analysis} \label{sec_ana}

\subsection{ACS photometry}\label{sec_phot}

Photometry is performed using the stellar-photometry package
DOLPHOT\footnote{http://purcell.as.arizona.edu/dolphot}, a modified
version of the HSTphot package to do photometry on \HST/WFPC2 images
\citep{dolp00}. DOLPHOT is equipped with a module for doing photometry
on ACS images which uses a look-up table to help determine the
spatially varying point spread function (PSF) at each detector
location.  We run DOLPHOT on the $B_{435}$, $R_{625}$ and H$\alpha$
images separately.  The filter-dependent model PSFs are used as
initial guesses for the actual PSFs of the images being analyzed; the
former are slightly adjusted using isolated stars to correct for
observation-specific focus and/or thermal conditions of the
telescope. For stars with a signal-to-noise ratio ($S/N$) lower than
10 aperture photometry is used. Source detection is run twice, with
sources that are detected in the first pass removed during the second
pass. Then, for each star, a photometric solution is iteratively
derived with the sky and neighbors subtracted until the photometry of
the star and its neighbors converge to a stable solution.  Sources
that are separated by fewer than 2 pixels are merged into a single
source.

DOLPHOT runs on the distorted ({\tt flt}) images to avoid photometry
errors associated with the pixel resampling that is necessary to
correct for the significant geometric distortion of ACS images. The
resulting output catalog, however, is aligned to a
distortion-corrected ({\tt drz}) image constructed with the STScI task
{\em multidrizzle}.  Optical source catalogs are derived for each
filter separately; these are combined using a match criterion of a
maximum separation of 1.5 pixels. DOLPHOT produces magnitudes for a
4-pixel aperture that are corrected for charge-transfer efficiency
degradation. We compute a filter-dependent aperture correction to 10
pixels using aperture photometry of a few isolated stars; these
corrections are between $-$0.08 and $-$0.10 mag. Together with the
aperture corrections from 10 pixels to ``infinity'' and the zero-point
offsets \citep{siriea05} they are applied to obtain calibrated
photometry in the STMAG system. The approximate detection limits for
$S/N = 5$ are $B_{435}\approx27$, $R_{625}\approx26.7$, and
$H\alpha\approx25.6$.

In an attempt to remove artifacts and non-stellar objects from the
source list, we generally only consider stars with $\chi^2 \leq 3$,
$S/N \geq5$, $-0.5\leq$ sharpness $\leq+0.5$, crowding $\leq$ 2 mag,
error flags $<8$ and object classifications appropriate for stellar
sources (type $\leq2$) in all three filters; see the DOLPHOT manual
for details. Even after applying these criteria, the final source list
is not guaranteed to be free of artifacts. For example, in the
$R_{625}$ versus $B_{435}-R_{625}$ color-magnitude diagrams (CMDs) a
small number ($\lesssim2$\% of the total number of sources detected in
all three bands) of faint, blue sources ($R_{625}>22.5,
B_{435}-R_{625}<2$) is prominent. Upon closer inspection, most turn
out to be associated with PSF artifacts near bright stars. For the
CMDs and color-color diagrams (CCDs) in this paper we have manually
cleaned this region of the CMD from artifacts, which explains the
sharp ``boundaries'' around this region in Figs.~\ref{fig_varext} and
\ref{fig_cmd} (left). For the identification of candidate counterparts to
{\em Chandra} sources, however, we visually inspect {\em all}
detections in the search area (\S\ref{sec_oid}) to check that no valid
optical source is erroneously eliminated and that no artifact is
mistaken for a real source.

We run DOLPHOT in the artificial-star mode to estimate the detection
completeness. For each of the magnitudes $R_{625}$=18, 19, ..., 25 we
create catalogs with 9000--9500 fake stars; $B_{435}$ and H$\alpha$
magnitudes are assigned according to the mean $B_{435}-R_{625}$ and
H$\alpha-R_{625}$ colors for that bin. The photometry is repeated in
each filter after adding each fake star in turn to the image. We then
check if the fake star is retrieved within 1 pixel of the input
position and 0.2 mag of the input magnitude. These tests were done for
one chip (1/8 of the field of view) only. We find that the
completeness is $\gtrsim$93\% for $R_{625}=22$ and drops to $\sim$27\%
for $R_{625}=25$.

\subsection{Improvement of the HST absolute astrometry} \label{sec_astro}

The coordinates in the optical catalogs are corrected for geometric
distortion but the {\em HST} absolute astrometry can be off by up to
3\arcsec~\citep[see][]{koekea05}. To improve the astrometric accuracy,
we tie the optical astrometry to the International Celestial Reference
System (ICRS) using stars in the UCAC2 catalog that have a positional
accuracy better than 0.070\arcsec~\citep{zachea04}. Since UCAC2 stars
are scarce and often saturated in the ACS images, we use a
ground-based $V$-band image of the LW to derive secondary astrometric
standards as an intermediate step. The ground-based image was obtained
with the CTIO-4m Mosaic camera (field of view 36\arcmin
$\times$36\arcmin) and reduced following the procedures described in
\cite{zhaogrindea05}.

We first derive an astrometric solution for a section of the CTIO
Mosaic image that includes the full ACS mosaic using 132 UCAC2
stars. Fitting for zero point, rotation angle, scale factor and
distortions results in a solution with rms residuals of
0.040\arcsec~in right ascension and 0.036\arcsec~in declination. Next,
the solution is transferred to one distortion-corrected image of each
ACS pointing using sets of $\sim$200 secondary standards selected from
stars that are unsaturated and relatively isolated in the Mosaic and
ACS images; we choose the stacked F658N images for this purpose which
should be clean of cosmic rays and other artifacts given the large
number of individual exposures. The resulting fits for zero point,
rotation angle and scale factor have rms residuals of, typically,
0.02\arcsec~in each coordinate. Finally, the solutions are transferred
to all ACS images of a given pointing using $\sim$2000 tertiary
standards; the associated errors are negligible. We estimate the final
1$\sigma$ accuracy with which the optical astrometry is tied to the
ICRS ($\sigma_o$) as the quadratic sum of the errors in the UCAC2
astrometry, the UCAC2--Mosaic tie and the Mosaic--{\em HST} tie,
i.e.~0.090\arcsec--0.095\arcsec. We adopt a uniform value of
0.095\arcsec\,for all pointings.

\subsection{Extinction} \label{sec_ext}

\begin{figure}
\centerline{\includegraphics[angle=-90,width=8.5cm]{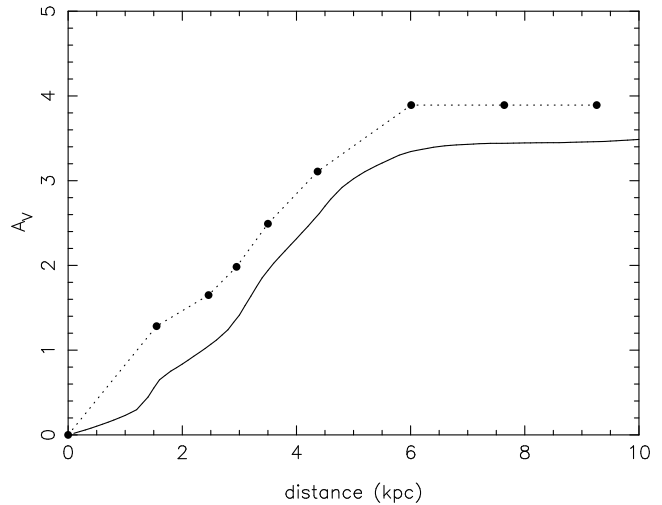}}
\caption{Extinction $A_V$ towards the LW as function of distance as
predicted by D03 (solid line) and M06 (dotted line). For the latter,
we have assumed $A_V = 8.3 \times A_{Ks}$. In our analysis, we adopt
the D03 curve (that extends out to a larger distance) after scaling it
to match the M06 curve at 8 kpc.}
\label{fig_ext}
\end{figure}

Extinction along the line of sight towards the LW is relatively low
for a bulge field but not entirely negligible, and needs to be
considered in order to derive intrinsic (unabsorbed) X-ray and optical
fluxes.  In Fig.~\ref{fig_ext} we compare two estimates from the
literature of the extinction as function of distance. The
three-dimensional $A_V$ maps by \cite{drimea03} (D03) are based on a
model for the Galactic dust distribution but are rescaled to match
COBE far-infrared data; the maps are thus limited to a resolution of
$\sim$0.35$^{\circ}$$\times$0.35$^{\circ}$. For the direction of the
LW, D03 predict that the bulk of absorbing material lies in front of 6
kpc after which $A_V$ levels off to 3.8; this includes a
$\sim$10\%-contribution from material that lies mainly beyond 10
kpc. However, for $|l| <20^{\circ}$ the underlying dust model is not
accurate. The $A_{Ks}$ extinction maps by \cite{marsea06} (M06),
available for $|l| <100^{\circ}$, $|b| < 10^{\circ}$ at
15\arcmin-resolution and computed out to $\sim$9 kpc, are derived by
comparing the colors of a simulated stellar population with the
corresponding observed 2MASS colors in a given direction.  The
predicted extinction towards the LW is somewhat higher than the value
from D03 ($A_V=3.9$ at 8 kpc for M06 versus $A_V=3.5$ for D03),
although both curves have similar shapes. For the total integrated
extinction $A_{V,{\rm max}}$, we will adopt the value from D03 that is
scaled to match $A_V$ from M06 at 8 kpc, i.e.~$A_{V,{\rm max}}=4.2$;
this corresponds to a neutral-hydrogen column density $n_{H,{\rm
max}}=7.6~10^{21}$ cm$^{-2}$ \citep{predschm95}. We use coefficients
from \cite{siriea05} to convert $A_V$ to extinction values in the {\em
HST} filters: $A_{B_{435}}=1.316\,A_V$, $A_{R_{625}}=0.851\,A_V$ and
$A_{{\rm H}\alpha}=0.815\,A_V$.

\begin{figure*}
\centerline{\includegraphics[angle=90,width=17cm]{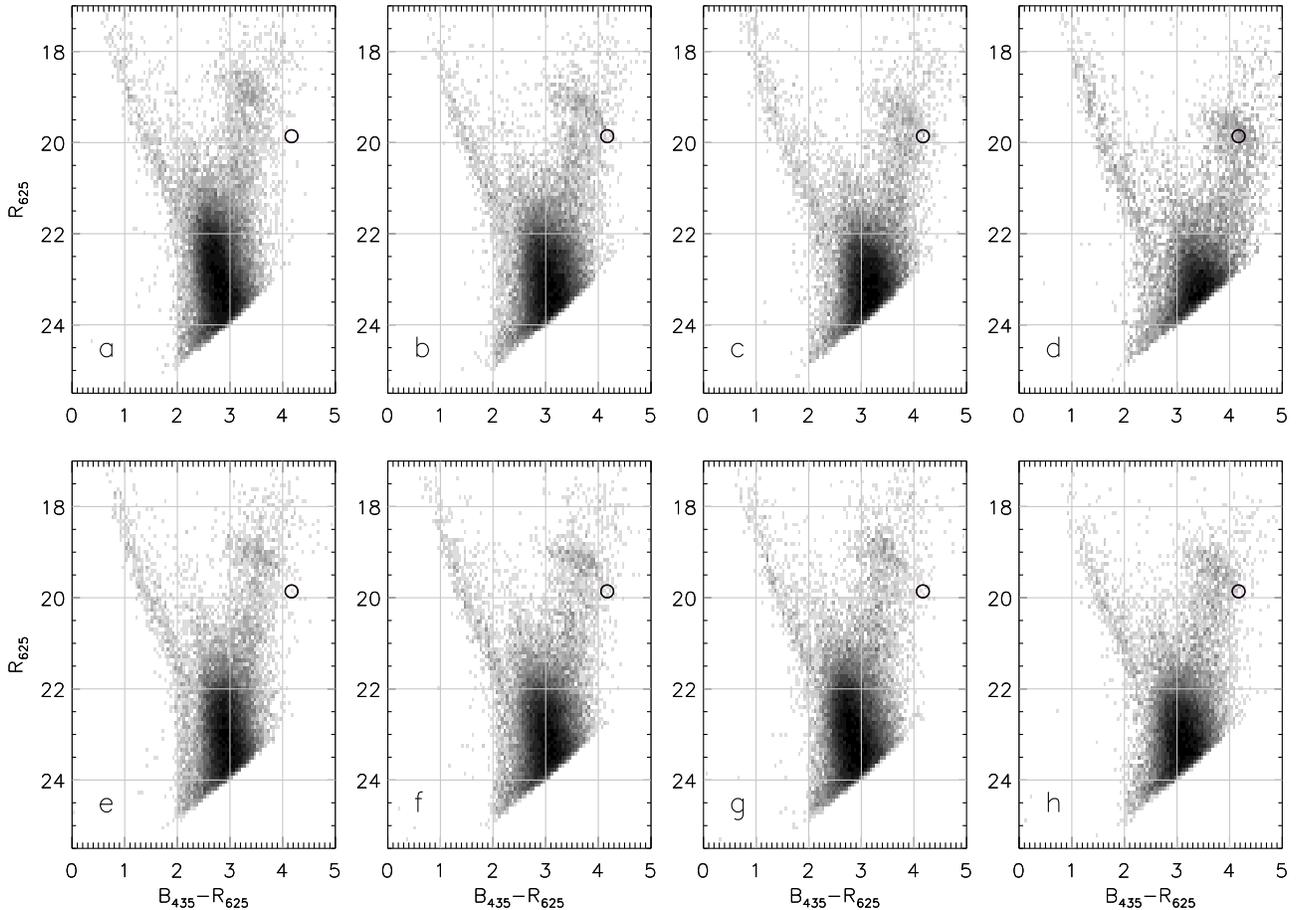}}
\caption{Color-magnitude diagrams extracted from each of the eight
chips of the ACS mosaic illustrate the variation of $A_V$ in our
field. The panels $a$ to $h$ are arranged according to their relative
location on the sky (so, panels $a$ and $h$ correspond to the
north-east and south-west corner of the mosaic, respectively). The
density of stars in each bin is indicated with a logarithmic gray
scale. The location of the red clump in panel $d$ at
$(B_{435}-R_{625},R_{625})\approx(4.2,19.9)$ is marked with a circle
in each panel. The variation of the observed locations of the
red-clump giants between panels, as well as the location of the bulk
of the stars, implies that the extinction $A_V$ varies by $\sim$30\%
across the field of the ACS mosaic.}
\label{fig_varext}
\end{figure*}

Besides the uncertainty in $A_{V,{\rm max}}$ and the distance
dependence of the extinction, spatial variations as function of
longitude and latitude also complicate matters. Owing to the weak age-
and metallicity-dependence of their luminosities and colors, red-clump
giants have frequently been used to trace extinction variations in the
bulge \citep[see for example][]{stanea97}. Fig.~\ref{fig_varext} shows
CMDs for each of the chips of the ACS mosaic (4 pointings, 2 chips per
pointing) with the red-clump giants visible in the upper right.  The
variation of their location implies that there are significant
variations in the extinction, especially in the east-west direction,
that amount to a maximum variation of $\sim$1 mag in $R_{625}$. This
corresponds to a $\sim$30\% variation in $A_V$ compared to $A_{V,{\rm
max}}=4.2$.

\subsection{X-ray spectral properties}

\begin{figure}
\centerline{\includegraphics[angle=0,width=8.5cm]{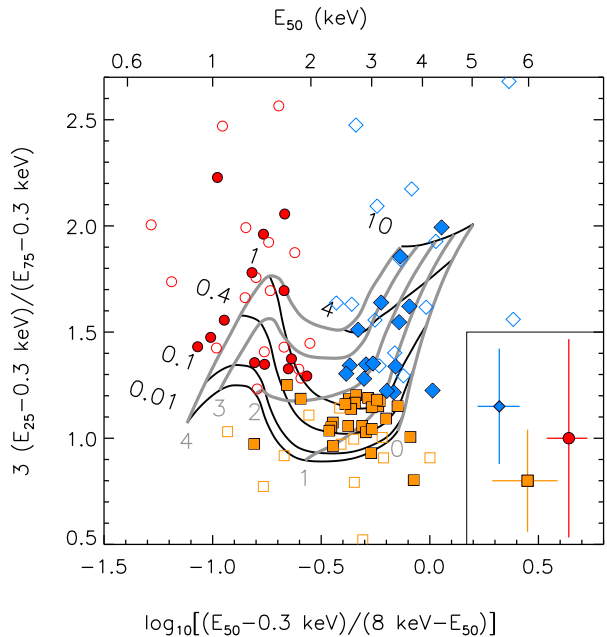}}
\caption{Quantile color-color diagram showing the {\em Chandra}
sources in the ACS field and their X-ray spectral classification. For
reference, we plot a grid for a power-law spectral model. The spectral
properties of a source can be derived from its location with respect
to the grid: black lines indicate contours of constant $n_H$ while the
power-law photon index $\Gamma$ is fixed along the thick gray
lines. The labels indicate the values of $\Gamma$ and $n_H$ (in units
of 10$^{22}$ cm$^{-2}$). Sources in the three spectral groups are
plotted with (red) circles (group 1; relatively soft sources),
(orange) squares (group 2; relatively hard sources) and (blue)
diamonds (group 3; absorbed sources). See Table~\ref{tab_ctr2flx} for
the group-specific spectral parameters. Sources with more and fewer
than 15 net counts ($B_x$) are plotted with filled and open symbols,
respectively. The median energy $E_{50}$ can be read off from the top
axis. The average errors for sources with $\geq$15 net counts are
shown in the inset on the bottom right for each spectral group. {\em
See the electronic edition of the journal for a color version.} }
\label{fig_qccd_all}
\end{figure}

All {\em Chandra} sources in the ACS field but one are detected with
fewer than 100 counts (the brightest source, our CV candidate LW\,1
(\S\ref{sec_col}), has 152 net counts); since this precludes the use
of spectral fitting, we resort to quantile analysis
\citep{hongschlea04} to derive the X-ray spectral properties.
In brief, quantile analysis constrains the spectral parameters and
column density from comparing the observed energy quantiles to those
expected for a spectral model of choice; in the {\em ChaMPlane}
analysis, we use the 25\%, 50\% and 75\% energy quantiles. Sources are
plotted, together with model grids, in a color-color plot from these
three quantiles as shown in Fig.~\ref{fig_qccd_all}.

Since measurement errors on the quantiles of individual sources
sometimes allow a broad range of parameters, we divide the sources in
three spectral groups and assign to all sources of a given group one
spectral model and $n_H$ based on the quantiles of their stacked
counts.  H09 describe in detail the process of assigning spectral
groups. The H09 spectral classification that we use in this paper is
represented in Fig.~\ref{fig_qccd_all}. For the soft sources (group
1), we assume that the spectra can be characterized by a 1-keV MeKaL
model which describes X-ray emission of an optically-thin hot plasma,
appropriate for the coronal emission of normal late-type stars; the
maximum $n_{H,{\rm max}}$ along the line of sight (\S\ref{sec_ext}) is
adopted. For the hard group (2) we use the spectral parameters from
H09 with a small adjustment, viz.~instead of the quantile-based $n_H$,
we also use $n_{H,{\rm max}}$ ($\sim$4\% difference). Parameters for
the absorbed group (3) are taken from H09 without adjustment. The
parameters for each spectral group are summarized in
Table~\ref{tab_ctr2flx}. To minimize the effect of an incorrect choice
of $n_H$ (not all sources are seen through the maximum column density
along the line of sight) we avoid using fluxes in the soft band. Since
not all sources are detected in the hard band ($H_c=2-8$ keV), we
compromise by working with broad-band ($B_x=0.3-8$ keV) fluxes.  For
the $B_x$ band, the countrate-to-(unabsorbed flux) conversion factor
is 2.2~10$^{-11}$ ergs cm$^{-2}$ count$^{-1}$ for group 2, while it is
16\% (46\%) higher for group 1 (3).

\begin{table}
\begin{center}
\caption{Spectral parameters for quantile spectral groups} \label{tab_ctr2flx}
\begin{tabular}{llll}
\hline
\hline
Group & Model     & Parameter     & $n_H$ \\ 
      &           &               & (10$^{22}$ atoms cm$^{-2}$) \\
\hline
1     & MeKaL     & $kT=1$ keV    & 0.76 \\ 
2     & power law & $\Gamma$=1.28 & 0.76 \\ 
3     & power law & $\Gamma$=1.21 & 1.95 \\ 
\hline
\end{tabular}
\end{center}
\end{table}

\subsection{Boresight correction and optical identification} \label{sec_oid}

We improve the absolute astrometry of the {\em Chandra} positions in
the ICRS (90\%-uncertainty is 0.6\arcsec) by identifying likely
matches between X-ray and optical sources. For this initial search for
counterparts we use the search area inside a combined (i.e.\
X-ray/optical) $\sim$95\% error radius that is the quadratic sum of
the 95\% X-ray error circles ($r_{95}$) and 2$\times$ the error on the
optical positions $\sigma_o$ (defined in
\S\ref{sec_astro})\footnote{Since $\sigma_o$ itself is the quadratic
sum of the errors in the directions of right ascension and
declination, 2 $\sigma_o$ represents a 98\% error radius if the errors
on the optical positions are distributed as a 2-dimensional
Gaussian. This means that our search area combines 95\% X-ray error
radii with 98\% optical error radii, where in practice the values for
the former are at least 1.7 larger than the latter.  Since the exact
shapes of the underlying distributions of the positional errors are
very difficult to measure, we choose not to refine this small
discrepancy in ``probability radii''.}. We accept an optical source as
a likely counterpart if it has unusual optical colors (blue, red
and/or with an H$\alpha$ excess with respect to the bulk of optical
sources with similar magnitudes) or if it has optical and X-ray colors
and an X-ray--to--optical flux ratio consistent with normal foreground
stars (only if there are no other contenders in the error circle). To
establish if the colors of a candidate counterpart are unusual, we
examine its location in a CMD that includes stars from just a small
(5\arcsec $\times$ 5\arcsec) region around the source since variations
in the extinction can make a star appear bluer or redder when the bulk
of the comparison stars are seen through a different extinction. We
also include matches with OGLE-III variables (A. Udalski, private
communication) that can be unambiguously identified in the {\em HST}
images and are unsaturated (saturation hampers the determination of
accurate optical positions); the properties of these variables are the
topic of a forthcoming paper.

After the first round of matching, an initial boresight is determined
from the weighted (by 1/$r_{95}^2$) average of the X-ray minus optical
positions of the likely matches. A uniform offset is determined for
all four ACS pointings since their astrometries have been put on the
same system; this is verified by examining the area where the
pointings overlap.  After correcting the X-ray positions for the
boresight, the matching is repeated after including in the search
radius a contribution from the boresight error, i.e.~we add a term (in
quadrature) that equals 2$\times$ the error on the boresight. The
resulting boresight between the original {\em Chandra} positions and
the corrected {\em HST} astrometry is almost negligible:
$\Delta\alpha=-0.02\arcsec\pm0.05$\arcsec,
$\Delta\delta=+0.05\arcsec\pm0.04$\arcsec\ (based on 11 matches).

\section{Results} \label{sec_res}

At least one astrometric optical match is found for each of the 100
{\em Chandra} sources.  Eleven sources have just 1 candidate
counterpart inside the search radius, while 10 sources have 2, 17 have
3, and 62 have 4 or more candidate counterparts. The average search
radius is 0.67\arcsec. In some cases, a bright star in or near the
error circle significantly reduces the sensitivity to detect
counterparts.  We mainly consider the optical colors of the
candidates, together with the ratio of the X-ray and optical fluxes
($f_{B_x}/f_{R_{625}}$ or $f_{X}/f_{R}$ in brief) and the X-ray
spectral properties to constrain which of them (if any) is the most
likely counterpart. We present cases that we have identified as likely
and possible CVs (or other accreting binaries) in
\S\ref{sec_xrb}. Additional identifications are briefly discussed in
\S\ref{sec_otherid}; the main purpose here is to compare the X-ray
properties of the candidate CVs with those of other source
classes. Constraints on the nature of the hard point sources in the LW
are discussed in \S\ref{sec_lnls}.

\subsection{Candidate accreting binaries} \label{sec_xrb}

\subsubsection{Selection based on optical colors} \label{sec_col}

\begin{table*}[t]
\begin{center}
\caption{Candidate accreting binaries in the LW selected on optical colors\label{tab_candid}}
\begin{tabular}{l@{\hspace{0.2cm}}c@{\hspace{0.2cm}}c@{\hspace{0.2cm}}ccccccccc@{\hspace{0.2cm}}l}
\hline
\hline
\multicolumn{1}{c}{(1)}  & (2)       & (3)      & (4)       & (5)   & (6)                 & (7)                 & (8)       & (9)       & (10)       & (11)             & (12)              & (13) \\
\multicolumn{1}{c}{ID}   & CXOPS\,J  & Counts   &  $E_{50}$ & Group & $\alpha_{\rm 2000}$ & $\delta_{\rm 2000}$ & $R_{625}$ & $B_{435}$ & H$\alpha$ & ${\rm Log} (f_X/f_R)$ & ${\rm Log} (f_X/f_R)_u$ & Comment \\
                         &           &          & (keV)     &       & ($^{{\rm h}}$ $^{{\rm m}}$ $^{{\rm s}}$) & ($^{\circ}$ ${\arcmin}$ ${\arcsec}$)               &           &           &           &                  &                    & \\
\hline
LW\,1  & 175120.9$-$293318 & 152$\pm$14      & 3.0$\pm$0.2    & 2 & 17 51 20.910 & $-$29 33 18.45 & 23.76  & 25.44  & 23.47  &    1.37(4) &    0.07(4) & \ldots \\ 
LW\,2  & 175118.1$-$293332 &   9$\pm$5       & 1.5$\pm$0.3    & 1 & 17 51 18.177 & $-$29 33 32.28 & 21.37  & 22.97  & 20.82  & $-$1.3(2)  & $-$2.1(2)  & dMe star?\\ 
LW\,3  & 175124.1$-$293738 &  14$\pm$6       & 1.8$\pm$0.3    & 2 & 17 51 24.123 & $-$29 37 39.00 & 22.30  & 24.06  & 22.07  &    0.53(6) & $-$0.78(6) & \ldots\\ 
LW\,4  & 175140.0$-$293557 &  76$\pm$10      & 2.9$\pm$0.3    & 1 & 17 51 40.019 & $-$29 35 57.60 & 23.81  & 25.03  & 23.36  &    0.0(2)  & $-$0.8(2)  & \ldots\\ 
LW\,5  & 175135.6$-$293754 & 19$\pm$6        & 3.3$\pm$0.7    & 2 & 17 51 35.679 & $-$29 37 55.51 & 23.82  & \ldots & 23.57  &    0.5(1)  & $-$0.8(1)  & \ldots\\ 
LW\,6  & 175122.8$-$293514 & 12$\pm$5        & 1.1$\pm$0.3    & 2 & 17 51 22.805 & $-$29 35 14.28 & 22.82  & 25.29  & 22.45  & $-$0.1(2)  & $-$1.4(2)  & \ldots\\ 
LW\,7  & 175111.8$-$293259 & 45$\pm$8        & 3.7$\pm$0.2    & 3 & 17 51 11.839 & $-$29 32 59.74 & 23.56  & 26.6   & 23.25  &    0.87(8) & $-$2.64(8) & \ldots\\ 
LW\,8  & 175118.7$-$293811 & 32$\pm$7        & 3.5$\pm$0.4    & 3 & 17 51 18.788 & $-$29 38 11.23 & 22.66  & 25.51  & 22.41  &    0.3(1)  & $-$3.2(1)  & \ldots\\ 
LW\,9  & 175118.5$-$293732 & 15$\pm$6        & 3.1$\pm$0.3    & 3 & 17 51 18.555 & $-$29 37 32.14 & 23.92  & 26.9   & 23.66  &    0.5(2)  & $-$3.0(2)  & \ldots\\ 
LW\,10 & 175115.8$-$293802 &  7$\pm$5        & 2.7$\pm$1.6    & 2 & 17 51 15.856 & $-$29 38 03.30 & 23.37  & 26.7   & 22.93  & $-$0.1(3)  & $-$1.4(3)  & \ldots\\ 
LW\,11 & 175134.8$-$293809 & 36$\pm$7        & 3.6$\pm$0.3    & 3 & 17 51 34.849 & $-$29 38 09.86 & 23.19  & 26.0   & 22.94  &    0.61(9) & $-$2.90(9) & \ldots\\ 
LW\,12 & 175141.4$-$293508 & 16$\pm$6        & 2.3$\pm$0.9    & 2 & 17 51 41.461 & $-$29 35 08.65 & 22.82  & 24.93  & 22.65  & $-$0.0(2)  & $-$1.3(2)  & \ldots\\ 
LW\,13 & 175142.8$-$293734 &  3$\pm$4        & \ldots         & 1 & 17 51 42.924 & $-$29 37 35.80 & 24.51  & 27.1   & 24.22  & $-$0.5(7)  & $-$1.3(7)  & \ldots\\ 
\hline
\end{tabular}
\end{center}
Properties of our best candidate CVs (LW\,1--4) and the more marginal
candidates (LW\,5--13). Column 1 gives the source name adopted in this
paper whereas column 2 gives the official {\em Chandra} source
name. Column 3 gives the net source counts in the $B_x$ band (0.3--8
keV), followed by the median energy $E_{50}$ in that band and the
quantile spectral group. Optical positions (columns 6 and 7) are 
given in units of hours, minutes, and seconds (for the right
ascension) and degrees, arcminutes, and arcseconds (for the
declination); errors in the absolute astrometry are
$\sim$0.095\arcsec. The remaining columns list the magnitudes, the
ratios of the observed ($\log (f_{X}/f_R)$) and intrinsic ($\log
(f_{X}/f_R)_u$) X-ray and optical fluxes, and comments. Fluxes are
corrected for extinction using the group-specific $n_H$ values listed
in Table~\ref{tab_ctr2flx}. Quoted errors in the flux ratios only
include a contribution from errors in the X-ray counts, not from
systematic errors. Errors are given in parentheses as the
(uncertainty) in the last digit.
\end{table*}

\begin{figure*}[thb]
\centerline{\includegraphics[width=17cm]{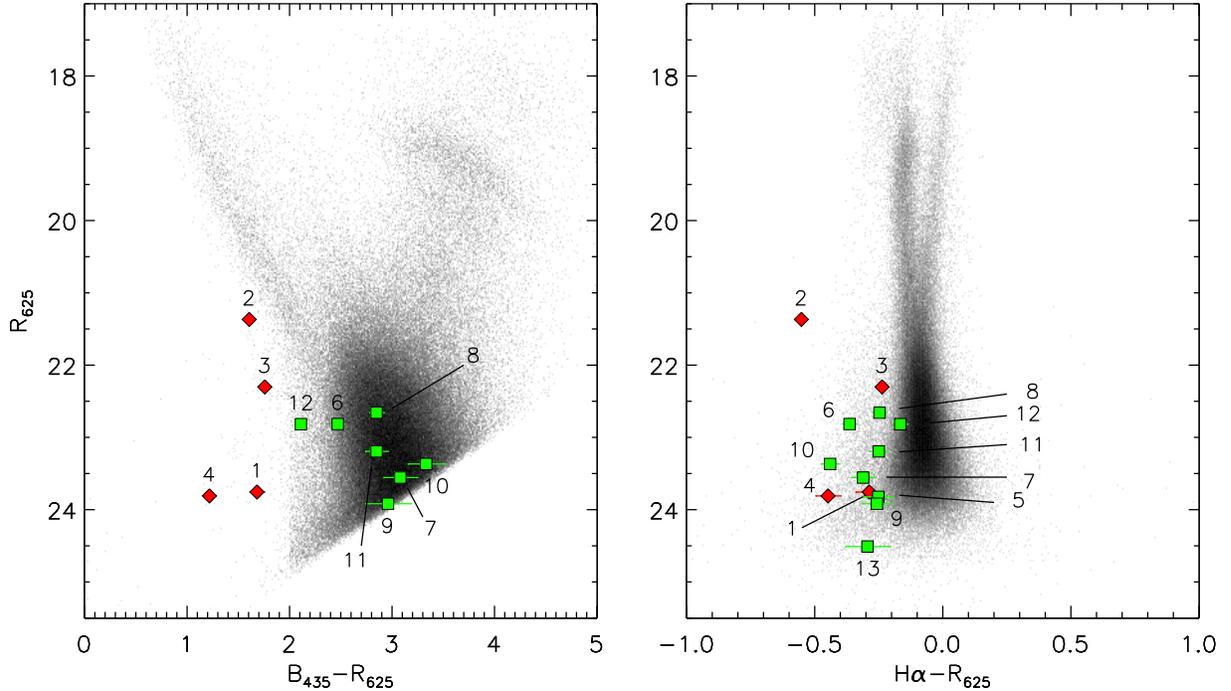}}
\caption{Color-magnitude diagrams with the candidate CVs from
Table~\ref{tab_candid} indicated with (red) diamonds (LW\,1--4; best
candidates) and (green) squares (LW\,5--13; possible CVs). Error bars
are plotted but in some cases are comparable to the size of the
symbols. Candidates for which $S/N<5$ in $B_{435}$ are not plotted in
the left panel. The density of stars in the CMDs is indicated with a
logarithmic gray scale; optical sources from all four ACS pointings
are included. {\em See the electronic edition of the journal for a
color version.}}
\label{fig_cmd}
\end{figure*}

\begin{figure*}[bht]
\centerline{\includegraphics[width=17cm]{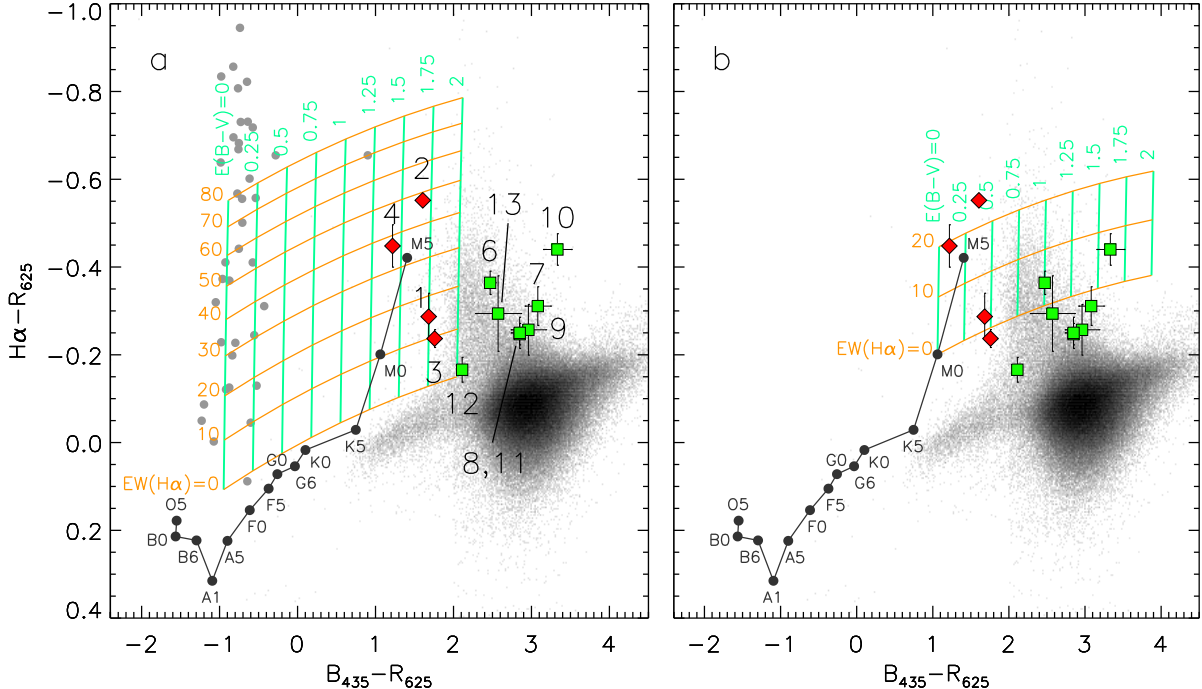}}
\caption{Color-color diagrams with the candidate CVs from
Table~\ref{tab_candid} marked with (red) diamonds (LW\,1--4; best
candidates) and (green) squares (LW\,5--13; possible CVs), and the
location of the unreddened main sequence (computed with spectral
standards from
\cite{jacoea84}) plotted as a dark-gray line. The grids show how the
synthetic colors for a disk spectrum ($a$) and an M0-dwarf spectrum
($b$) vary as function of reddening $E(B-V)$ (in steps of 0.25 mag)
and the equivalent width of H$\alpha$ (in steps of 10 \AA).  Grids
were computed with the {\tt synphot} package; we assumed that the
H$\alpha$ emission line has a Gaussian profile with a full width at
half maximum of 25 \AA~for the disk model, and 1.5 \AA~for the
M0-dwarf model. The small light-gray circles in panel $a$ are the
colors of SDSS CVs at $|b|>55^{\circ}$ and illustrate the (spread in
the) colors of real CVs. The location of optical sources detected in
all four ACS pointings is shown with a logarithmic gray scale. {\em
See the electronic edition of the journal for a color
version.}}\label{fig_ccd}
\end{figure*}

\begin{figure*}
\centerline{\includegraphics[width=15cm]{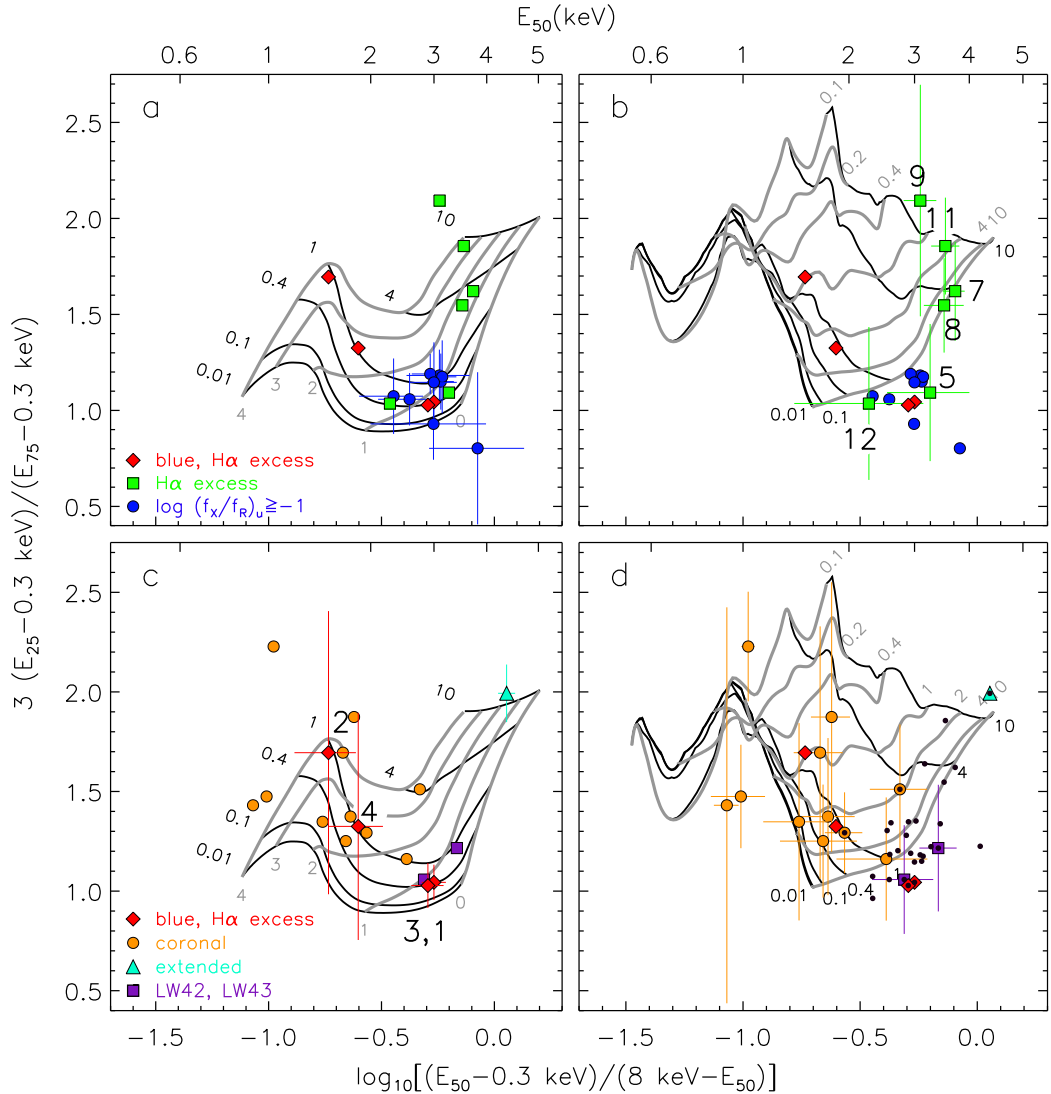}}
\caption{Quantile color-color diagrams with our best candidate CVs
LW\,1--4 marked in each panel with (red) diamonds and labelled in
panel $c$ only. The model grids on the left ($a,c$) are computed for a
power-law spectrum, while the grids on the right ($b,d$) are for a
single-temperature MeKaL model. In panels $a$ and $b$, we also include
the other candidate CVs: (green) squares are the sources with
H$\alpha$-excess fluxes (labelled in $b$) and (blue) circles the
sources with $\log (f_{X}/f_R)_u \gtrsim -1$. In panels $c$ and $d$,
we include the identifications with coronal sources ((orange) circles)
that are distinctly softer than most candidate CVs from $a$ and $b$.
Other noteworthy sources like the candidate AGN LW\,41 ((cyan)
triangle) and LW\,43 ((purple) square) are also included in $c$ and
$d$. Error bars are plotted in one panel only for clarity. Thick gray
lines are contours of constant temperature $kT$ or photon index
$\Gamma$; black lines are contours of constant $n_H$. Except for LW\,2
and 3, sources with $<$15 net counts are omitted. The small black
circles in panel $d$ mark sources with $S/N_{H_c} > 3$ that are
discussed in
\S\ref{sec_lnls}. {\em See the electronic edition of the journal for a color version.}}
\label{fig_qccd}
\end{figure*}

\begin{figure*}
\centerline{\includegraphics[width=16cm]{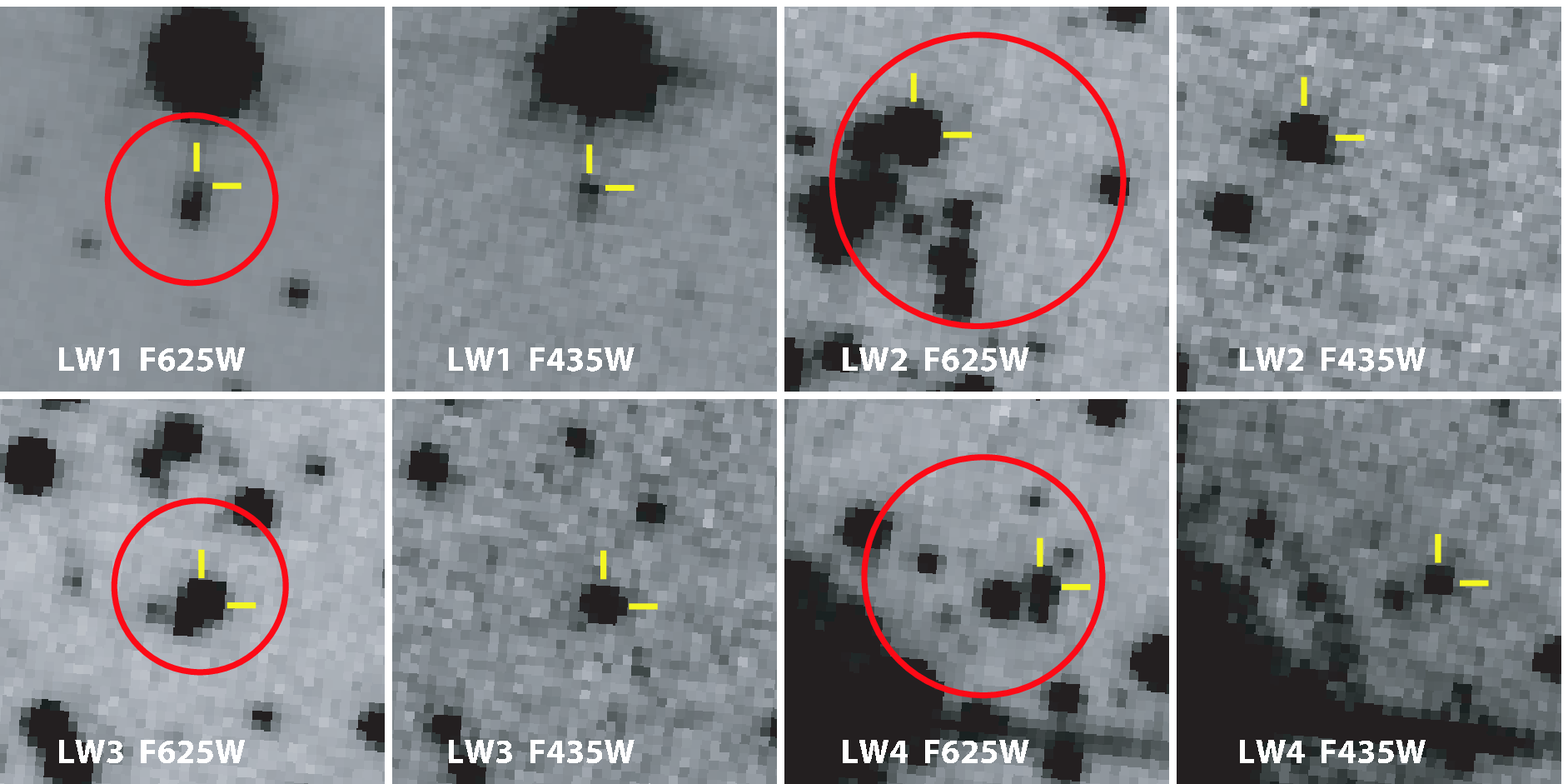}}
\caption{Finding charts in F625W and F435W for the four best candidate
CVs. In the F625W images, we show the area inside which we search
for optical counterparts with a (red) circle (with radius
0.44\arcsec~for LW\,1, 0.76\arcsec~for LW\,2, 0.45\arcsec~for LW\,3
and 0.62\arcsec for LW\,4). The candidate CVs are indicated with
(yellow) tick marks. The scaling of the images is adjusted to
highlight the relatively blue color of the proposed
counterparts. North is up, east is left. {\em See the electronic
edition of the journal for a color version.}
\label{fig_fc}}
\end{figure*}

Most known CVs have blue colors that arise from the accretion disk,
the accretion stream or sometimes the white dwarf, and show Balmer
emission lines from the disk or stream.  Therefore, if a {\em Chandra}
source can be matched with a blue optical source that also has excess
H$\alpha$ flux, the source is a good candidate for being a CV.  Based
on this criterion, we initially find four good candidates, LW\,1 to
LW\,4, whose properties are summarized in
Table~\ref{tab_candid}. Figs.~\ref{fig_cmd} and \ref{fig_ccd} show
their locations in the $R_{625}$ versus $B_{435}$$-$$R_{625}$ and
$R_{625}$ versus H$\alpha$$-$$R_{625}$ CMDs and $B_{435}$$-$$R_{625}$
versus H$\alpha$$-$$R_{625}$ CCD ((red) diamonds); for comparison, we
also include the other optical sources in all four ACS pointings.  In
the LW, however, where we observe sources at various distances and
thus through a range of foreground extinctions, the interpretation of
optical colors is not straightforward: a distant CV can be confused
with a nearby late-type star.  To demonstrate this, first we compare
the expected colors for CVs to the colors of LW\,1 to LW\,4. Using the
{\tt synphot} package, we have simulated the colors for a spectrum
with a flux-wavelength dependence $F \propto
\lambda^{-2/3}$, i.e.~appropriate for the emission from an accretion
disk \citep{warn95}. The grid in Fig.~\ref{fig_ccd}a shows how these
synthetic colors behave as a function of reddening $E(B-V)$ and
equivalent width (EW) of the H$\alpha$ line. For this particular
spectrum and the maximum reddening towards the LW ($E(B-V)_{\rm
max}=A_{V,{\rm max}}/3.1\approx1.35$), no CVs are expected with
$B_{435}-R_{625}\gtrsim1.2$. However, small-scale variations in the
extinction or different assumptions for the underlying optical
spectrum (for example, an additional contribution from the light of
the secondary) can explain redder systems like LW\,1 and LW\,3. To
illustrate the spread in intrinsic colors of real CVs, we have plotted
in Fig.~\ref{fig_ccd}a with small gray circles the colors of
CVs\footnote{Colors were computed using {\tt synphot} and the SDSS
spectra from Data Release 7. A small fraction (3.2\% of the total
area) of the F435W filter bandpass is not covered by the spectra,
which means that the actual $B_{435}-R_{625}$ colors of the SDSS CVs
are slightly bluer than plotted in Fig.~\ref{fig_ccd}a.} from the
Sloan Digital Sky Survey (SDSS; \cite{szkoea02} and follow-up papers)
that lie at $|b| \ge 55^{\circ}$ and are therefore not or moderately
reddened (the total Galactic extinction in the direction of these
systems is only $0.02 \lesssim A_V \lesssim 0.13$; D03). The location
of the unreddened main sequence (gray line) shows how our color
selection of CV candidates can cause confusion with nearby M
dwarfs---also common X-ray sources. The color grid for an M0 dwarf in
Fig.~\ref{fig_ccd}b shows this in more detail; for example, the
optical colors of LW\,4 could point at a CV with $E(B-V)\approx1.4$
and EW(H$\alpha$) $\approx$ 30--40 \AA, but can also be explained by
an M2--3 dwarf with $E(B-V)\approx0$ and EW(H$\alpha$) $\approx$ 10
\AA. LW\,4 is unlikely to be of later spectral type since the source
is too blue, or of earlier type which would require that EW(H$\alpha$)
$\gtrsim$ 15 \AA. In the spectroscopic study of X-ray--selected
M-dwarfs by \citet{mochea02}, EW(H$\alpha$) does not exceed 11
\AA~among the 54 stars in the sample.

We try to resolve this ambiguity by considering other properties, too.
For LW\,1, LW\,3 and LW\,4 the ratio of the intrinsic X-ray to optical
flux, $\log (f_X/f_R)_u$, exceeds $-$1 if we assume that they are seen
through $n_{H,{\rm max}}=7.6~10^{21}$ cm$^{-2}$
(Table~\ref{tab_candid}, column 12), and is even higher ($>0$) if they
are in fact nearby and therefore unabsorbed (Table~\ref{tab_candid},
column 11; see \S\ref{sec_fxfo} for details on estimating
$(f_X/f_R)_u$). These inferred values of $\log (f_X/f_R)_u$ are high
compared to the typical range for active M dwarfs (see
\S~\ref{sec_fxfo}).  For LW\,1 and LW\,3, their classification as CVs
is also the more plausible one on the basis of their X-ray colors; the
quantile diagrams of Fig.~\ref{fig_qccd} shows that their spectra are
hard ($kT>10$ keV or $\Gamma\lesssim1$) which is unusual for stellar
coronal sources.  Also for LW\,4, which is softer in X-rays, we favor
the explanation as CV: if the star is indeed an unabsorbed M2 dwarf,
the absolute $V$ magnitude ($M_V=9.9-10.2$; \cite{carrostl96,bess91}),
and $V-R_{625}$ color ($\sim$0.84 as derived with {\tt synphot} using
a spectral standard) would put this star at a distance of
$\sim$7.2--6.3 kpc, which is inconsistent with the implied low
reddening from Fig.~\ref{fig_ccd}b.  The case of LW\,2 is less
clear. If the source is an M5 dwarf with EW(H$\alpha$) $\approx$ 10
\AA~(see Fig.~\ref{fig_ccd}b), its distance would be $\sim$250--700 pc
($M_V=12.3-14.7$; $V-R_{625}\approx0.8$) and thus $E(B-V)\approx0.1$
which is roughly consistent with the estimated $E(B-V)$ from the
color-color grids (Fig.~\ref{fig_ccd}b). Moreover, $\log
(f_X/f_R)_u<-1$, and the soft X-ray colors are more typical for
coronal than accreting sources. On the other hand, dwarf novae in
outburst can have soft X-ray spectra \citep[see for
example][]{wheaea96}. Such a state is often accompanied by weak Balmer
emission lines or even absorption lines as the accretion disk becomes
optically thick. We cannot determine if LW\,2 also displayed this
behavior. The X-ray photons from LW\,2 mainly come from the second
ACIS exposure taken 3 days after the {\em HST} observation (the source
was not detected separately in the other two ACIS exposures).  We
conclude that of our four blue and H$\alpha$-excess sources, LW\,1,
LW\,3 and LW\,4 are likely CVs while LW\,2 is a possible CV or a dMe
star.  Fig.~\ref{fig_fc} shows their finding charts.

Nine additional {\em Chandra} sources are matched with
H$\alpha$-excess sources. These are marked with (green) squares in
Figs.~\ref{fig_cmd}, \ref{fig_ccd} and \ref{fig_qccd}ab and included
in Table~\ref{tab_candid} as LW\,5 to LW\,13. We note that LW\,6 and
LW\,12 have a relatively bright neighbor
($R_{625}\approx17$) at a separation of 0.6\arcsec~and
0.5\arcsec~($\sim$13 and 10 pixels), respectively, which may have
affected their photometry. On the other hand, their PSFs look clearly
separated and the photometric quality flags do not signal any
problems. We consider these nine sources as less secure CV candidates
as LW\,1, 3 and 4 because most are not or only slightly blue compared
to the bulk of the optical sources; others have $S/N<5$ in $B_{435}$
or are not detected in $B_{435}$ at all.
While this is perhaps more suggestive of them being late-type stars
rather than CVs, we still mention them for two reasons. First, some
CVs do not appear blue in $B-R$ (the dwarf nova identified with the
{\em Chandra} source CX\,1 in the globular cluster M\,55 is an
example; see
\cite{bassea08}). And second, the X-ray colors (and for LW\,5 also the
$(f_{X}/f_R)_u$ ratio) of some of these sources are too high for
normal or coronally-active stars, see Figs.~\ref{fig_qccd}a and b. Not
much can be said about the X-ray colors of LW\,6, LW\,10 and LW\,13
that are detected with fewer than 15 net counts, while the spectra of
LW\,7, LW\,8, LW\,9 and LW\,11 appear absorbed with
$n_H\gtrsim10^{22}$ cm$^{-2}$ ($A_V\gtrsim5.6$).  This requires some
source of extra extinction, for example originating inside a binary
system or provided by the host galaxy of an active galactic nucleus
(AGN).  In the latter case the H$\alpha$ excess could be explained in
the relatively unlikely event that an emission line happens to be
redshifted in the F658N band. For coronal sources such high $n_H$
values are not expected. Occasionally, the column density is enhanced
when a star undergoes a large coronal flare but these cases are rare
and not yet fully understood \citep{gued04}.

\subsubsection{Selection based on $f_{X}/f_R$ ratios} \label{sec_fxfo}

\begin{table}
\begin{center}
\caption{Candidate accreting binaries in the LW selected on
$(f_{X}/f_R)_u$\label{tab_candfxfo}}
\begin{tabular}{l@{\hspace{0.2cm}}ll@{\hspace{0.2cm}}l@{\hspace{0.18cm}}r@{\hspace{0.18cm}}r}
\hline
\hline
ID     & CXOPS\,J          & Counts    &  $E_{50}$ & ${\rm Log} (f_{X}/f_R)$   & ${\rm Log} (f_{X}/f_R)_u$ \\ 
       &                   &           & (keV)     & (minimum)             & (minimum)            \\
\hline
LW\,14  &175127.9$-$293508  & 14$\pm$5       & 2.4(6)   & 0.9(2) &  $-$0.4(2)    \\ 
LW\,15  &175119.1$-$293740  & 41$\pm$8       & 3.1(6)   & 1.66(9) &     0.36(9)  \\ 
LW\,16  &175127.0$-$293706  & 46$\pm$8       & 2.6(5)   & 1.37(8) &     0.06(8)  \\ 
LW\,17  &175130.0$-$293613  & 88$\pm$11      & 2.9(3)   & 0.94(5) &  $-$0.37(5)  \\ 
LW\,18  &175114.2$-$293224  & 22$\pm$7       & 3(1)     & 1.1(1) &  $-$0.2(1)    \\ 
LW\,19  &175133.6$-$293313  & 77$\pm$10      & 3.1(2)   & 0.50(6) &  $-$0.81(6)  \\ 
LW\,20  &175130.7$-$293232  & 54$\pm$9       & 3.2(3)   & 0.65(7) &  $-$0.65(7)  \\ 
LW\,21  &175120.7$-$293417  & 18$\pm$6       & 3.8(9)   & 0.4(1) &  $-$0.9(1)    \\ 
LW\,22  &175124.0$-$293532  & 13$\pm$5       & 1.2(7)   & 0.6(2) &  $-$0.7(2)    \\ 
LW\,23  &175139.6$-$293811  & 34$\pm$7       & 2.3(5)   & 0.5(1) &  $-$0.8(1)    \\ 
LW\,24  &175128.9$-$293412  & 14$\pm$6       & 2.7(7)   & 0.6(2) &  $-$0.7(2)    \\ 
LW\,25  &175123.5$-$293755  & 38$\pm$8       & 3.0(4)   & 0.3(9) &  $-$1.0(9)    \\ 
\hline
\end{tabular}
\end{center}
For each source we give the name adopted in this paper, followed by
the official source name, the net counts in the $B_x$ band, the median
energy in $B_x$, and the ratio of X-ray and optical fluxes computed
for the brightest optical source in the error circle
(observed: $\log (f_{X}/f_R)$; unabsorbed: $\log (f_{X}/f_R)_u$). All
sources have quantile spectral group 2; optical and X-ray fluxes are
therefore dereddened with $n_H=7.6~10^{21}$ cm$^{-2}$. Quoted errors
in the flux ratios only include a contribution from errors in the
X-ray counts, not from systematic errors. Errors are given in
parentheses as the (uncertainty) in the last digit.
\end{table}

It is well-known \cite[e.g.][]{stocea91,krauea99} that the
distribution of the ratio of intrinsic X-ray to optical fluxes
$(f_X/f_{{\rm opt}})_u$ for accreting binaries extends to
significantly higher values than for coronal sources, whereas it
largely overlaps with the range observed for galaxies. The CVs in
\cite{verbea97} have $\log (f_X/f_V)_u$ between $-$2.4 and +0.5 where
we converted from their 0.5--2.5 keV band to our $B_x$ band using the
2-keV thermal-bremsstrahlung model used in \cite{verbea97}.  On the
other hand, for the nearby stars in \cite{schmlief04} and the active
binaries in \cite{dempea} and \cite{dempea97}, $\log
(f_x/f_V$)$_{u,{\rm max}}\approx-1$ in the ROSAT band which, for the
X-ray spectra typical for coronal sources ($kT\approx1$ keV), would
give a similar value in our $B_x$ band. \cite{krauea99} find $<\log
(f_X/f_V)_u>=-1.33\pm0.50$ for dMe stars (that have the highest flux
ratio's among single stars) with only a few outliers that have $\log
(f_X/f_V)_u>-0.5$.  Therefore, even in the absence of any optical
matches with distinct colors, we can identify CV (or AGN) candidates
among the {\em Chandra} sources by considering the minimum value of
$(f_{X}/f_R)_u$ that results from the brightest optical source in the
search area. The difficulty of applying this diagnostic to our sample
is that we only have a rough estimate of the extinction that is needed
to convert between observed and intrinsic fluxes.  We adopt the
quantile-group specific $n_H$ values (Table~\ref{tab_ctr2flx}) but
stress that this is only an estimate. For example, if the fluxes for a
source in group 1 are corrected for $n_H=7.6~10^{21}$ cm$^{-2}$
whereas in reality $n_H=0$, $\log (f_X/f_R)_u$ would be underestimated
by 0.8 dex.  Similarly, the $\log (f_X/f_R)_u$ value for a group-2
source could be up to 1.3 dex too small, although the situation is
likely not that serious as their distribution indeed appears to be
concentrated between $n_H\approx4~10^{21}$ and 10$^{22}$ cm$^{-2}$ in
Fig.~\ref{fig_qccd_all}. For group-3 sources, for which both $n_H$ and
the underlying spectra are poorly constrained, $\log (f_X/f_R)_u$ is
even more difficult to estimate. We further assume that
$f_V$$\approx$$f_{R}$, to facilitate comparison with the
literature. We convert our optical magnitudes to fluxes using the zero
points from \cite{siriea05} and apply a small correction factor to
specify fluxes in a 1000-\AA--wide band width\footnote{To compute the
optical flux $f_R$, we multiply the flux density by the filter width
for which we adopt the often-used value of 1000 \AA\ instead of
$\sim$415 \AA, which is the actual width the F625W filter. The
correction factor that we add to the measured $\log (f_X/f_R)$ and
$\log (f_X/f_R)_u$ ratios is therefore $\log({\frac{1000}{415}})=0.38$
.}, also to make comparison with published values easier.

Based on the above ranges, we select possible CVs as sources with
$\log (f_{X}/f_R)_{u,{\rm min}} \gtrsim -1$. The twelve sources
(besides LW\,1, 3, 4 and 5) that satisfy this criterion are listed in
Table~\ref{tab_candfxfo} and marked in the quantile diagrams of
Fig.~\ref{fig_qccd}a and b ((blue) circles). The relatively hard
spectra support their classification as non-coronal sources. However,
both their quantile and $(f_{X}/f_R)_u$ values prevent the distinction
from AGN, and the colors of the astrometric optical matches do not put
additional constraints on the source classifications. Possibly, the
true optical counterparts are too faint to be detected. The detection
limit of $R_{625}\approx26.7$ implies that any undetected counterparts
have $\log (f_{X}/f_{R})_{u,{\rm min}}=0.2-0.9$ for the sources in
Table~\ref{tab_candfxfo}, which would be consistent with accreting
binaries.

\subsection{Other identifications} \label{sec_otherid}

We defer a detailed discussion of stellar coronal sources in the LW to
a follow-up paper, but for the purpose of putting the X-ray properties
of our candidate CVs into a broader context we also briefly discuss
likely coronal sources here. Two sources are matched with bright UCAC2
stars. Our low-resolution optical follow-up spectra show them to be a
K and F star. Since these stars are heavily saturated in the ACS
images, it is impossible to look for other optical sources in the
error radius. In the field of view of the ACS images, stars of this
brightness are however rare and chance alignments are unlikely.  Other
likely coronal sources are found among astrometric matches of {\em
Chandra} sources to OGLE-III variables (A. Udalski, private
communication); while we have not done a test yet to estimate how many
of the thirteen OGLE matches could be spurious, a detailed analysis of
matches between the {\em Chandra} sources in the two other Windows
with OGLE-II variables \citep{vdbergea06} showed that $<$10\% are
expected to be random. Also, twelve of the light curves vary
periodically and suggest binary periods from 0.36 days to 34
days. This could explain the X-rays as being generated by
tidally-induced rapid rotation.  The UCAC2 stars (LW\,26 and 27) and
the variables (LW\,28--40) are plotted with (orange) circles in
Figs.~\ref{fig_qccd}c and d for sources with $>$15 net counts. They
are predominantly soft in X-rays but we note the relatively hard X-ray
colors of LW\,32 and LW\,37. The derived $n_H$ values are consistent
with $n_{H,{\rm max}}$ or lower, suggesting many may be foreground
coronal sources. They occupy a distinctly different region of the
quantile diagram than most candidate CVs from \S\ref{sec_xrb}
(Figs.~\ref{fig_qccd}a and b). The properties of LW\,26--40 are
summarized in Table~\ref{tab_otherid}. For the {\em Chandra} sources
that are matched to OGLE variables we list the optical ACS source that
lies closest to the OGLE position\footnote{after correcting for a
small systematic offset between the OGLE and ACS coordinate systems,
computed using four unambiguous identifications}. In three cases, two
optical sources lie at a similar distance and we give the coordinates,
magnitudes and flux ratios of both. The relatively high $(f_X/f_R)_u$
ratios for the fainter of the two candidate identifications to LW\,29
and LW\,36 make them the less likely of the two candidate matches to
these {\em Chandra} sources.

\begin{figure}
\centerline{\includegraphics[width=8.5cm]{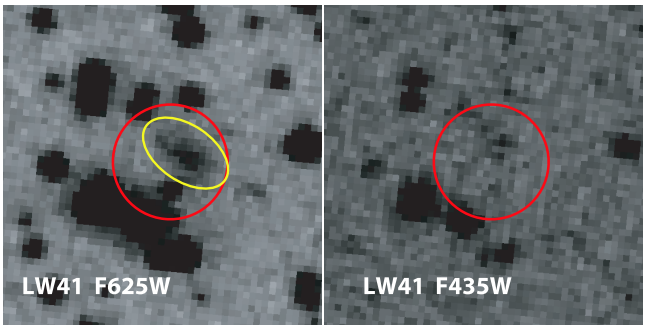}}
\caption{Finding charts in F625W and F435W of the region around
LW\,41. An extended and red object (indicated by (yellow) ellipse)
lies near the center of the area inside which we search for
optical counterparts marked with the (red) circle (radius
0.45\arcsec). North is up, east is left. {\em See the electronic
edition of the journal for a color version.}}
\label{fig_galaxy}
\end{figure}

Another interesting identification is the red extended object that
matches with LW\,41, see Fig.~\ref{fig_galaxy} and
Table~\ref{tab_otherid}. Its position in the quantile diagram suggests
that it is highly absorbed and that the spectrum is consistent with a
power law with photon index 1--2 ((cyan) triangle in
Figs.~\ref{fig_qccd}c and d). We suggest that this source is a type-II
AGN, and that some sources in this region of the quantile diagram are
too.

Finally, we draw attention to two relatively hard sources, LW\,42 and
LW\,43, that both are matched to optical sources with remarkable
colors but that cannot readily be classified ((purple) squares in
Figs.~\ref{fig_qccd}c and d; Table~\ref{tab_otherid}). One of the
candidate counterparts to LW\,42 is detected in the $B_{435}$ band but
in neither the $R_{625}$ nor the H$\alpha$ images, and must therefore
be very blue. The $n_H$, and therefore $\log (f_X/f_B)_u$, is not
well-constrained. The source was assigned quantile-group 3 which gives
$\log (f_X/f_B)_u \approx -3.5$. However, if we assume that the
underlying spectral model is a power law, lower $n_H$ values are also
allowed; for $n_H=7.6~10^{21}$ cm$^{-2}$, $\log (f_X/f_B)_u
\approx 0.1$. The X-ray spectrum is too hard to be from a stellar
coronal source.
LW\,43 has an exceptionally red optical source in the error circle
which looks normal in H$\alpha-R_{625}$. The spectrum is harder than
expected for a coronal source, and we suggest an obscured AGN is the
most plausible explanation.

The remaining 57 LW sources that are not listed in
Tables~\ref{tab_candid}--\ref{tab_otherid} do not have candidate
counterparts with unusual colors, nor are the limits on their
$(f_X/f_R)_u$ values stringent enough to exclude that they are normal
or active stars. For sources with sufficient counts, the only
distinction that can be made is between probable stellar coronal and
non-coronal sources based on X-ray spectral hardness that (as
Fig.~\ref{fig_qccd} shows) are distinctly different for these two
source classes.

\begin{table*}
\begin{center}
\caption{Other identifications of {\em Chandra} sources in the LW\label{tab_otherid}}
{\footnotesize
\begin{tabular}{l@{\hspace{0.2cm}}c@{\hspace{0.2cm}}c@{\hspace{0.2cm}}c@{\hspace{0.2cm}}c@{\hspace{0.2cm}}ccccccc@{\hspace{0.2cm}}c}
\hline
\hline
\multicolumn{1}{c}{(1)}   & (2)               & (3)      & (4)       & (5)  & (6)                  & (7)                 & (8)       & (9)       & (10)       & (11)             & (12)              & (13)    \\
\multicolumn{1}{c}{ID}    & CXOPS\,J          & Counts   &  $E_{50}$ & Group & $\alpha_{\rm 2000}$ & $\delta_{\rm 2000}$ & $R_{625}$ & $B_{435}$ & H$\alpha$ & ${\rm Log} (f_X/f_R)$ & ${\rm Log} (f_X/f_R)_u$ & Comment \\
                          &                   &          & (keV)     &       & ($^{{\rm h}}$ $^{{\rm m}}$ $^{{\rm s}}$) & ($^{\circ}$ ${\arcmin}$ ${\arcsec}$)  &           &           &           &                  &                    &    \\
\hline
LW\,26     & 175130.6$-$293227 & 38$\pm$8      & 0.91$\pm$0.07  & 1 & 17 51 30.598 & $-$29 32 27.82 & 11.8$^a$ & \ldots & \ldots &  $-$4.37(9)$^a$ & $-$5.21(9)$^a$ & UCAC2 \\ 
LW\,27     & 175128.1$-$293703 & 15$\pm$6      & 1.0$\pm$0.2    & 1 & 17 51 28.133 & $-$29 37 03.37 & 12.2$^a$ & \ldots & \ldots &  $-$4.6(2)$^a$  & $-$5.5(2)$^a$  & UCAC2 \\ 
LW\,28     & 175137.5$-$293602 & 29$\pm$7      & 1.03$\pm$0.04  & 1 & 17 51 37.533 & $-$29 36 02.58 & \ldots   & \ldots & 15.52  &     $-$3.0(1)$^c$ & $-$3.9(1)$^c$    & OGLE  \\ 
LW\,29$^b$ & 175129.5$-$293602 & 16$\pm$6      & 1.7$\pm$0.3    & 1 & 17 51 29.555 & $-$29 36 02.45 & 23.70    & 26.8   & 23.42  &     $-$0.1(2)     & $-$0.9(2)        & OGLE  \\ 
\ldots     & \ldots            & \ldots        & \ldots    & \ldots & 17 51 29.548 & $-$29 36 02.55 & 21.05    & 25.06  & 20.86  &     $-$1.2(2)     & $-$2.0(2)        & OGLE  \\ 
LW\,30     & 175118.9$-$293733 & 14$\pm$6      & 1.7$\pm$0.3    & 1 & 17 51 18.905 & $-$29 37 33.65 & 21.05    & 24.81  & 20.86  &     $-$1.2(2)     & $-$2.0(2)        & OGLE  \\ 
LW\,31     & 175139.7$-$293546 & 9$\pm$5       & 1.3$\pm$0.4    & 1 & 17 51 39.765 & $-$29 35 47.03 & \ldots   & 17.75  & 16.63  &     $-$3.0(2)$^c$ & $-$3.8(2)$^c$    & OGLE  \\ 
LW\,32     & 175136.4$-$293747 & 21$\pm$6      & 2.8$\pm$0.5    & 3 & 17 51 36.400 & $-$29 37 47.50 & 18.51    & 22.04  & 18.34  &     $-$1.5(1)     & $-$5.0(1)        & OGLE  \\ 
LW\,33     & 175132.5$-$293636 & 46$\pm$8      & 1.9$\pm$0.2    & 1 & 17 51 32.528 & $-$29 36 36.35 & 19.61    & 22.70  & 19.44  &     $-$1.27(8)    & $-$2.07(8)       & OGLE  \\ 
LW\,34     & 175119.1$-$293551 & 17$\pm$6      & 1.7$\pm$0.3    & 1 & 17 51 19.138 & $-$29 35 51.52 & 20.07    & 23.73  & 19.95  &     $-$1.3(1)     & $-$2.1(1)        & OGLE  \\ 
LW\,35     & 175137.3$-$293409 & 15$\pm$6      & 1.8$\pm$0.2    & 1 & 17 51 37.307 & $-$29 34 09.67 & 20.14    & 23.61  & 19.96  &     $-$1.6(2)     & $-$2.4(2)        & OGLE  \\ 
LW\,36$^b$ & 175131.0$-$293216 & 17$\pm$6      & 1.7$\pm$0.4    & 2 & 17 51 31.065 & $-$29 32 16.98 & 19.13    & 22.69  & 18.98  &     $-$1.5(2)     & $-$2.8(2)        & OGLE  \\ 
\ldots     & \ldots            & \ldots        & \ldots    & \ldots & 17 51 31.065 & $-$29 32 17.16 & 23.82    & \ldots & 23.59  &      0.4(2)     & $-$0.9(2)        & OGLE  \\ 
LW\,37$^b$ & 175114.5$-$293756 & 18$\pm$6      & 2.5$\pm$0.7    & 2 & 17 51 14.579 & $-$29 37 56.18 & 22.65    & 25.64  & 22.49  &      0.0(2)     & $-$1.3(2)        & OGLE  \\ 
\ldots     & \ldots            & \ldots        & \ldots    & \ldots & 17 51 14.565 & $-$29 37 56.04 & 20.36    & 24.33  & 20.13  &     $-$0.9(2)     & $-$2.2(2)        & OGLE  \\ 
LW\,38     & 175133.2$-$293718 & 14$\pm$5      & 1.5$\pm$0.2    & 1 & 17 51 33.242 & $-$29 37 18.59 & 19.43    & 21.73  & 19.30  &     $-$1.9(2)     & $-$2.7(2)        & OGLE  \\ 
LW\,39     & 175138.8$-$293409 & 18$\pm$6      & 1.4$\pm$0.3    & 1 & 17 51 38.886 & $-$29 34 09.20 & 20.72    & 23.76  & 20.54  &     $-$1.2(2)     & $-$2.0(2)        & OGLE  \\ 
LW\,40     & 175138.3$-$293308 & 14$\pm$6      & 1.0$\pm$0.2    & 1 & 17 51 38.351 & $-$29 33 09.18 & 18.08    & 19.18  & 18.14  &     $-$2.4(2)     & $-$3.2(2)        & OGLE  \\ 
LW\,41     & 175137.4$-$293515 & 61$\pm$9      & 4.4$\pm$0.2    & 3 & 17 51 37.485 & $-$29 35 15.29 & 22.50    & \ldots & 25.0   &      0.56(7)    &   $-$2.95(7)     & extended \\ 
LW\,42     & 175133.3$-$293430 & 30$\pm$7      & 3.4$\pm$0.3    & 3 & 17 51 33.317 & $-$29 34 30.63 & \ldots   & 27.1   & \ldots &      2.1(1)$^d$     &  $-$3.5(1)$^d$       & blue \\ 
LW\,43     & 175127.8$-$293510 & 28$\pm$7      & 2.8$\pm$0.5    & 2 & 17 51 27.893 & $-$29 35 10.48 & 21.92   & 26.6  & 21.79  &     $-$0.1(1)     &  $-$1.4(1)       & red \\ 
\hline										       
\end{tabular}
}
\end{center}
Properties of the identifications discussed in
\S\ref{sec_otherid}. See Table~\ref{tab_candid} for an explanation of
the columns.

$^a$LW\,26 and LW\,27 are saturated in our ACS images. In column 8 we give the UCAC2
    magnitude from \cite{zachea04} that is between $V$ and $R$, and use it to
    compute the X-ray--to--optical flux ratios.\\
$^b$The identification of the OGLE variable in the ACS image is ambiguous, therefore we list
    both candidate optical counterparts.\\
$^c$The optical source is saturated in the F625W image; we assume $R_{625}$=H$\alpha$ to compute the X-ray--to--optical flux ratios.\\
$^d$The optical source is not detected in the F625W and H$\alpha$ images; we use $B_{435}$ to compute the X-ray--to--optical flux ratios.

\end{table*}

\subsection{Constraints on the hard (2--8 keV) source population} \label{sec_lnls}

H09 use the number-density versus flux distribution ($\log{N} -
\log{S}$) of {\em Chandra} sources in seven inner-bulge fields
including the three Windows to study the hard point-source population
and its radial distribution in the central region of the Galaxy. Their
results suggest that the high concentration of hard sources around
SgrA* is part of a population that extends out to at least the LW
where the hard-source density is still significantly elevated above
the extra-galactic $\log{N} - \log{S}$ curve, in contrast to what is
seen in Galactic-plane fields at larger offsets from SgrA*. Of the 109
LW sources with $S/N_{H_c}>3$ that are used in the analysis by H09, 29
are covered by the ACS mosaic\footnote{One can look up the IDs of
these sources in Table~\ref{tab_srclist}.}. In Fig.~\ref{fig_qccd}d
these are marked with black dots. The average flux limit for a
$S/N_{H_c}=3$ detection in the field of the ACS mosaic is 2~10$^{-15}$
ergs s$^{-1}$ cm$^{-2}$ (2--8 keV) for a power-law spectrum with
photon index $\Gamma=1.7$ absorbed by $n_{H,{\rm max}}=7.6~10^{21}$
cm$^{-2}$. Based on this limit and the high-latitude hard-source
density from \cite{kimea07}, we expect that $\sim$17$\pm$4 (Poisson
errors only) hard sources are galaxies. We now consider what
constraints we can set on the nature of the remaining, i.e.~Galactic,
sources based on our optical identifications.

Of the 29 sources, we have plausible identifications for six: two are
identified with periodic OGLE variables and likely coronal sources
(LW\,32 and LW\,33; these are among the ``softest'' hard sources), one
is the extended object and probable AGN LW\,41, one is the red object
and candidate AGN LW\,43, and two are our likely CVs LW\,1 and
LW\,3. The hard X-ray spectra of the latter two could indicate they
are magnetic systems that often have harder X-rays than seen in CVs of
other types \citep[e.g.][]{heinea08}.  The properties of eleven
sources appear to exclude a coronal nature: one is the blue source
LW\,42, seven sources have high values for $\log (f_X/f_R)_{u,{\rm
min}}$ (LW\,15--17, LW\,19, LW\,20, LW\,23, LW\,25), and three sources
that are matched with H$\alpha$-excess objects (LW\,7, LW\,8, LW\,11)
are also unlikely to be coronal as they lie in quantile group 3,
i.e.~show indications of extra absorption in X-rays. As outlined in
\S\ref{sec_xrb} and \S\ref{sec_otherid}, these sources can be either
accreting binaries or AGN. We have no obvious optical clues to the
nature of the remaining twelve sources that all have at least one
astrometric match with non-distinct colors.  Since they lie to the
right of most coronal sources in Figs.~\ref{fig_qccd}c and d, they are
not likely to be normal or active stars.  The true counterparts may be
below the detection limit in each or at least one filter which
prevents us from measuring colors. In summary, the only constraint we
can set on the Galactic hard sources is that they are a mix of stellar
coronal and accreting sources but are, not unexpectedly, dominated by
the latter. We come back to nature of the hard-source population in
\S\ref{sec_gcxrb}.

\section{Discussion} \label{sec_dis}

Our search for CVs in the LW has resulted in three likely CVs (LW\,1,
LW\,3, LW\,4) and 22 candidate CVs (LW2, LW\,5--25). In what follows
we set constraints on the distances of the best candidates
(\S\ref{sec_dist}), and discuss the CV space density in the bulge
(\S\ref{sec_density}). We consider our identifications in the context
of hard X-ray sources in the central bulge in
\S\ref{sec_gcxrb}.

\subsection{Distance constraints for the CV candidates} \label{sec_dist}
With the limited information we have on our candidate CVs, it is
difficult to meaningfully constrain their distances. We adopt two
methods. One approach is to estimate minimum distances based on the
$R_{625}$ and H$\alpha$ magnitudes, using the empirical relation
between the equivalent width of the H$\beta$ emission line,
EW(H$\beta$), and the absolute $V$ magnitude of the accretion disk:
EW(H$\beta$)=0.3\,$M_V^2$+exp(0.55\,($M_V$--4)) for ${\rm EW(H}\beta)
\gtrsim 15$ \AA, and $M_V\lesssim+6$ for weaker lines \citep{patt84}.
A relatively safe upper limit for EW(H$\beta$) is
EW(H$\alpha$). Comparing the spectroscopically-determined equivalent
widths of H$\alpha$ and H$\beta$ in the SDSS CVs (\cite{szkoea02} and
follow-up papers), we find that the mean ratio
EW(H$\beta$)/EW(H$\alpha$) is 0.58 with a standard deviation 0.38 in a
sample of 197 spectra.  CV candidates LW\,1, 3 and 4 have
EW(H$\alpha$) values between $\sim$10 \AA~and $\sim$38 \AA~as inferred
from their locations with respect to the grid in
Fig.~\ref{fig_ccd}a. Assuming ${\rm EW(H}\beta)\leq {\rm
EW(H}\alpha)$, it follows that $M_V\lesssim 6$ for LW\,1 and LW\,3,
and $M_V\lesssim 8.8$ for LW\,4. Taking into account the $A_V$ curve
from D03, we find that $d_{{\rm min}}$ is between 3.5 and 6.2 kpc if
$V-R_{625}=0$ or between 4.2 and 9.3 kpc if $V-R_{625}=1$ although,
given the scatter of $\pm$$\sim$1.5 around the ${\rm EW(H}\beta)$ --
$M_V$ relation, LW\,4 can be as close as 2.7 kpc.  This suggests our
candidates are much farther away than the majority of known CVs in the
field. There are a few caveats however. We have assumed that the
systems are not internally absorbed; if internal absorption is
present, we have overestimated the actual distances. Furthermore, our
estimate for EW(H$\alpha$) relies on the choice of model for the
optical spectrum for which we have assumed a power-law model.  If
instead the white-dwarf spectrum---typically with broad Balmer
absorption lines---contributes significantly around H$\alpha$, our
derived EW(H$\alpha$) will be underestimated (and the minimum
distances overestimated). Also, the ${\rm EW(H}\beta)-M_V$ relation
applies to systems with disks, and is not appropriate for magnetic
systems in which no disks or truncated disks are formed.

Therefore, we also use another approach to constrain the distances of
our candidate CVs, viz.~we test the hypothesis that they are at
similar distances than most known CVs. The compilation of CVs by
\cite{akea08} shows that the majority of known CVs is closer than
$\sim$2 kpc,
therefore we ask: based on the local CV density, how many CVs do we
expect in the volume limited by a distance $d=2$ kpc and the field of
view of the ACS mosaic?  To answer this question, we integrate the CV
density $n_{{\rm CV}}$ along the line of sight and assume that it
varies as a function of distance according to the Galaxy model by
\cite{picarobi04} (see next section for more details). For a maximum
distance of 2 kpc the model is dominated by the Galactic disk whereas
the bulge is not relevant.  The resulting number of CVs, regardless of
the selection effects that these systems would be subjected to in
order to be detected, is 0.18 and 0.54 for a local CV density
1~10$^{-5}$ and 3~10$^{-5}$ pc$^{-3}$, respectively (these values
cover the range of densities found in recent studies, see e.g.
\cite{akea08}, \cite{grinhongea05},
\cite{pretea07}).
The probability to detect 3 or more nearby CVs when in fact $\sim$0.54
are expected is low ($\lesssim$1\% for a Poisson
distribution). Therefore we conclude that we have probed the CV
population at larger distances.

\subsection{The CV density towards the bulge} \label{sec_density}

Since we can derive only crude distances for our candidate CVs, it is
not possible to use them to compute the CV density towards the LW
directly. Instead, we assume that outside the local neighborhood the
ratio of the CV-to-stellar space density is the same as locally, and
compare the number of CVs we expect to detect 
to the number we actually see.

We first calculate the number of stars in a volume enclosed by the
6.6\arcmin$\times$6.6\arcmin~field of the ACS mosaic and the maximum
distance $d_{{\rm max}}$ at which we can detect the brightest CVs (38
kpc for $M_V$=+3, \cite{patt98}). We adopt the Galaxy model by
\cite{picarobi04} which consists of two parts. The first is a
multi-component exponential disk with a radial scale length of 2.35
kpc, vertical scale heights between 60 and 185 pc (depending on the
age of the component) and a central hole with a scale length of 1.31
kpc. The second component is a triaxial bulge with a major axis of
1.82 kpc that is rotated with respect to the line connecting us and
the Galactic Center by $\sim$11$^{\circ}$, and axis ratios
1:0.31:0.26. The central nuclear bulge is not included, but is not
relevant here given the separation of the LW from the Galactic
Center. \cite{picarobi04} assume that the distance to the Galactic
Center is 8.5 kpc.  The star densities are normalized such that the
disk gives a local density of 0.145 stars pc$^{-3}$ and that the bulge
has 11.5 stars pc$^{-3}$ in the center of the Galaxy.  See
\cite{picarobi04} for details of this model (sech$^2$ density profile
and ``Pad7.9'' luminosity function).  Integrating the star density
along the line of sight gives a total of about 12~10$^{6}$ stars.
Assuming that throughout this volume the ratio of CVs to stars is the
same as locally implies that there are $\sim$800--2400 CVs in this
volume (for a local CV density of 1~10$^{-5}$ and 3$~10^{-5}$
pc$^{-3}$).

We verify how well the number of stars predicted compares to the
number observed. To account for the distance out to which stars of
different masses can be detected, we fold the Galaxy model with the
local luminosity function from \cite{bahcsone80}.  We calculate the
number of stars predicted between $V=22$ and 24. By choosing this
range, we avoid sampling the steep-sloped part of the luminosity
function (that applies for bright stars) at bulge distances where the
star density is highest. This would have the adverse effect of making
the predicted number very sensitive to the assumptions for $A_V$. We
neglect differences between the luminosity functions in $V$ and
$R_{625}$ which we believe are modest: the stars that dominate the
optical detections are F to K stars which have $V-R<1$. Furthermore,
we use the D03 $A_V$ curve, again scaled to match M06 at 8 kpc
(\S\ref{sec_ext}). We take into account the detection completeness
derived from the DOLPHOT simulations, which for the magnitude range
chosen is $\sim$92\% on average. We find that about 3.8 times more
stars are predicted than we observe in the $R_{625}$ images
($\sim$10$^5$ versus 2.6~10$^4$). For an increase (decrease) of $A_V$
by 50\%, this factor is 3.1 (4.5). For now we do not explore in detail
the origin(s) of this discrepancy that could be due to the Galaxy
model, differences between the local and bulge luminosity functions,
or uncertainties in $A_V$.  We simply reduce the number of expected
CVs by this factor, which gives $\sim$210--630 CVs expected for the
$A_V$ adopted (260--790 and 180--540 if $A_V$ is 50\% higher or
lower).

We simulate our detection efficiency for CVs in a similar way as
\cite{koenea08}.
We create a sample of 10$^{6}$ fake CVs to which we randomly assign an
X-ray luminosity $L_X$ and $f_{X}/f_R \approx f_{X}/f_V$ ratio based
on the observed distributions by \cite{verbea97}, and a distance
somewhere between us and $d_{{\rm max}}$.  We assume the CV-density
profile follows the stellar-density profile by \cite{picarobi04} but
scale it to match the local CV density. We make sure that the
combination of distance, $L_X$ and $f_X/f_V$ does not give an absolute
optical magnitude outside the observed range for CVs, $+3\lesssim M_V
\lesssim +12$ \citep{patt98}.  For each fake CV we then assess whether
it falls above the detection limit in the {\em Chandra} data (for a
10-keV thermal-bremsstrahlung model and a flux limit corresponding to
15 net counts ($B_x$) or $S/N_{B_x}\approx 3$) and in the ACS images
(we use $R_{625,{\rm lim}}=24$, slightly fainter than LW\,1 and LW\,4;
for fainter stars the detection completeness drops below $\sim$30\%),
after taking into account extinction.  The resulting detection
efficiency is about 17\%. This number does not depend much on the
choice of X-ray spectrum (a 2-keV thermal bremsstrahlung model gives a
similar result), but is sensitive to $A_V$: an increase in $A_V$ by
50\% reduces the efficiency to 6\%, while a 50\% decrease gives an
efficiency of 32\%. The efficiency is further reduced by a factor
0.7--0.8 (depending on $B_{435}-R_{625}$ color) to $\sim$13\% (4.4\%
or 24\% if $A_V$ is 50\% higher or lower) after taking into account
the three-band detection completeness of the optical photometry. The
simulated efficiency still overestimates the real efficiency. For
example, a fraction of CVs have no or weak H$\alpha$ emission lines
(e.g.~dwarf novae in outburst) and would consequently not be selected
as candidate CVs.

In summary, the predicted number of CVs is at most
$\sim$0.13$\times$210=27 to $\sim$0.13$\times$630=80 for the nominal
$A_V$ (12--35 or 43--130 if $A_V$ is 50\% higher or lower) where the
lower and upper ends of the ranges are for a local CV density
1~10$^{-5}$ and 3~10$^{-5}$ pc$^{-3}$. This should be compared to
three likely CVs and 21 possible CVs detected with net counts
consistent with 15 (within 2$\sigma$) or more. At face value, this
implies that the ratio of the CV and stellar densities in the bulge is
similar to or lower than the local value. If this ratio is indeed
between 6.9~10$^{-5}$ and 2.1~10$^{-5}$, it suggests a local CV
density closer to 1~10$^{-5}$ pc$^{-3}$ (corresponding to the lower of
the two ratios) than to 3~10$^{-5} $pc$^{-3}$. In that case, at 8.2
kpc a peak in the CV density towards the LW is reached of
$\sim$7~10$^{-4}$ pc$^{-3}$; averaged over the bulge (7--10 kpc) the
density would be 6.2~10$^{-4}$ pc$^{-3}$. In the extreme case that
none of our 21 possible CVs turn out to be true CVs, the relative CV
density decreases towards the bulge, and/or we have severely
overestimated our detection efficiency, and/or $A_V$ is higher than
our nominal value (which we know is the case in at least part of our
field). Given the uncertainties (in the $L_X$ and $f_X/f_V$
distributions, $A_V$, X-ray and optical spectra) and the small sample,
we cannot draw firm conclusions. In a follow-up paper where we report
on the search for candidate CVs in the other two Windows, the
constraints on the bulge CV density will be reconsidered with improved
statistics.

\subsection{The hard X-ray sources in the central bulge} \label{sec_gcxrb}

One of our main goals is to investigate the nature of the hard (2--8
keV) source population in the central bulge that culminates in the
high concentration of sources around SgrA*. H09 found that the
dominating source classes in the SgrA* field and in the LW have
similar X-ray properties, and could be part of a single bulge
population whose projected number density is roughly inversely
proportional to the angular offset from the Galactic Center. This
implies that the suggestion by \cite{munoarabea04}, viz.~that the
majority of hard sources around SgrA* are magnetic CVs in the bulge
(specifically: intermediate polars), should also hold true for the LW.
We have identified two hard LW sources, LW\,1 and LW\,3, as plausible
magnetic CVs. Below we investigate if all hard Galactic sources in the
ACS mosaic ($\sim$10$\pm$4 in total, i.e.~the number of sources with
$S/N_{H_c}$$>$3 minus the contribution from AGN and the two coronal
sources; \S\ref{sec_lnls}) can be accounted for by bulge CVs.

For a spatial distribution that follows the normal star density and is
subject to the selection effects described in \S\ref{sec_density}, it
is expected that about two-thirds of all observed CVs lie between
7--10 kpc. Let us assume that LW\,1 and LW\,3 indeed lie in the
bulge. For convenience we put them at the Galactic-Center distance
according to the \cite{picarobi04} model, i.e.~at 8.5 kpc, which means
their intrinsic absolute magnitudes are $M_{R_{625}}\approx 5.8$ and
4.3. They would represent only part of the total bulge CV population,
since at 8.5 kpc only CVs with $M_{R_{625}} \lesssim +6$ can be
detected (for $R_{625,{\rm lim}}$=24). Based on the time-averaged
$M_V$ distribution of CVs by \cite{patt98}, the fraction of CVs with
$M_V \lesssim +6$ is $\sim$27\% (for $V-R_{625}=0$) or $\sim$41\% (for
$V-R_{625}=1$). Other references give lower fractions. From our
compilation of 31 CVs and CV candidates in globular and open clusters
(47\,Tuc, NGC\,6397, M\,67, NGC\,6791) we find $\sim$16\% and
$\sim$29\% for the two choices of $V-R_{625}$ color, respectively. Out
of 24 nearby CVs with astrometric parallaxes
\citep{thor03,thorea08,pattea08} 1 (4\%) has $M_V \lesssim +6$ while 3
(12\%) have $M_V \lesssim +7$, where we have assumed $A_V=0$ and have
taken the observed $V$ magnitude from \cite{rittkolb03}. While clearly
the $M_V$ distribution is uncertain, these fractions are consistent
with having optically identified only 2 out of 10$\pm$4 (14--33\%)
candidate bulge CVs, and therefore allow all hard, non-coronal sources
in the ACS mosaic to be accounted for by bulge CVs (we assume that for
CVs $L_X$ and $M_V$ are not correlated). This would imply an {\em
observed} hard-CV density in the bulge of 0.74--1.7~10$^{-5}$
pc$^{-3}$, assuming these sources lie between 7 and 10 kpc. To make
this number match with the estimated total CV density in the bulge of
6.2~10$^{-4}$ pc$^{-3}$ (\S\ref{sec_density}), hard CVs must
constitute between 1--3\% of all bulge CVs. Despite uncertainties in
the $L_X$ and X-ray--spectral distributions of CVs, we make a rough
estimate of how realistic this fraction is. Our hard sources have $L_X
\gtrsim 4.2~10^{31}$ ergs s$^{-1}$ (2--8 keV) or $L_X \gtrsim
1.3~10^{31}$ ergs s$^{-1}$ (0.5--2 keV) for the X-ray spectra
assumed. Comparing this to the ROSAT CVs in \cite{verbea97} (0.5--2.5
keV), we estimate that we have observed the brightest $\sim$30\% of
CVs; therefore 4--9\% of all CVs should be hard. To put this number in
context: $\sim$13\% and $\sim$10\% of the CVs in the compilation by
\cite{rittkolb03} are polars and intermediate polars, respectively,
which typically have the hardest X-ray spectra among the known
classes. On the other hand, \cite{liebea03} suggest a lower fraction
of magnetic CVs, about~$\sim$10\%. In summary, according to our rough
estimate, magnetic bulge CVs or IPs specifically offer a viable
explanation for the hard Galactic sources in the LW if the CV-to-star
ratio is constant.

LW\,1 and LW\,3 could be the first members of the inner-bulge
hard-source population to be optically identified. If most hard
sources near SgrA* indeed have the same nature as LW\,1 and LW\,3,
there is no chance of finding their optical or infrared counterparts
due to the much higher extinction towards the Galactic Center. We note
that not one of the hard LW sources matches with a bright
($R_{625}<17$) optical source, which implies that in the LW they are
not dominated by wind-fed Be HMXBs. Their bright companions cannot be
missed: a typical B0V secondary with $M_V=-4$ and $V-R
\approx -0.1$, seen through $A_V=4.2$, would have $R_{625}\approx
14.7$ at 8.5 kpc and $R_{625}
\approx 16.3$ at twice that distance. The same conclusion was reached
for the hard sources around SgrA* \citep{laycea05}.

\section{Conclusions} \label{sec_con}

We have searched for CVs in a low-extinction bulge field at a
1.4$^{\circ}$ angular separation from the Galactic Center. We
discovered three likely CVs and 22 additional candidates that could
also be coronally-active stars or background galaxies. Distance
estimates place these systems at $\gtrsim$2 kpc. We investigate the
CVs space density towards the bulge and find no large discrepancy
between the number of observed and expected CVs in the direction of
our field if the CV-to-star ratio is fixed at the local value. In that
case, a local CV density of 1~10$^{-5}$ pc$^{-3}$ is preferred by our
findings. However, low statistics and uncertainties in the extinction
and X-ray and optical properties of CVs prevent us from drawing firm
conclusions. We will address the issue of the bulge CV density in more
detail in similar studies of other low-extinction fields.

We identify two hard X-ray sources as plausible CVs. Their X-ray
properties are similar to those of the bulk of the {\em Chandra}
sources around SgrA*. If the Galactic-Center hard (2--8 keV) X-ray
source population extends out to the field studied in this paper, as
suggested by H09, these two systems could be the first optically
identified members of this elusive population. We suggest that the
other hard Galactic sources in our field can also be explained by CVs.
This makes the sources presented in this paper prime targets for
further study.  Ground-based optical spectroscopy will be impossible
for candidates like LW\,1, but we encourage follow-up work on the more
isolated ones. With the $\sim$900 ks {\em Chandra} exposure of our
field described in \cite{revnea09}, the X-ray spectral and variability
properties of the sources can be studied in more detail allowing one
to place better constraints on their nature.

\begin{acknowledgements}
The authors wish to thank A.~Udalski for providing a list of OGLE-III
variables in our field prior to publication. We are grateful to
A. Dolphin for help with running the Dolphot photometry package. Our
research made use of the Sloan Digital Sky Survey database. This work
was supported in part by STScI/HST grant HST-GO-10353.01 and
NASA/Chandra grants GO6-7088X, GO7-8090X and GO8-9093X. Results in
this paper are based on observations made with the NASA/ESA Hubble
Space Telescope, obtained at the Space Telescope Institute which is
operated by the Association of Universities for Research in Astronomy,
Inc. under the NASA contract NAS 5-26555.
\end{acknowledgements}

{\it Facilities:} \facility{CXO (ACIS)}, \facility{HST (ACS/WFC)}

\bibliographystyle{apj}

\end{document}